\documentclass[a4paper,11pt]{article}
\usepackage{jheppub}
\usepackage[utf8]{inputenc}

\usepackage{graphicx} % Required for inserting images

\usepackage{amsxtra}
\usepackage{float}
\usepackage[dvipsnames]{xcolor}
\usepackage{tikz-cd}
\usepackage{macros}
\usepackage{mdframed}

\usepackage{dcolumn}
\newcolumntype{L}{D{.}{.}{1.1}}

\tikzset{gaugebl/.style={circle,draw=black,fill=black,inner sep=2.5pt}}
\tikzset{hasse/.style={circle, fill,inner sep=2pt}}

\definecolor{goodgreen}{RGB}{55,169,49}

\tikzset{gauge/.style={inner sep=1mm,draw=none,fill=white,minimum size=2mm,circle, draw}}
\tikzset{flavour/.style={draw=none,minimum size=0.3mm,fill=white, regular polygon,regular polygon sides=4,draw}}
\tikzset{hasse/.style={circle, fill,inner sep=2pt}}

\usepackage{todonotes}

\preprint{Imperial/TP/26/AH/04}

\title{A Tale of Two Orbits: Non-Simply Laced Mirror
}

\author[a]{ Amihay Hanany,}
\author[a]{ Deshuo Liu$\,$}

\affiliation[a]{Theoretical Physics Group, The Blackett Laboratory, Imperial College London,\\ Prince Consort Road
London, SW7 2AZ, UK}

\emailAdd{a.hanany@imperial.ac.uk}
\emailAdd{deshuo.liu21@imperial.ac.uk}

\abstract{
A three-dimensional $\mathcal{N}=4$ gauge theory is constructed whose Higgs branch realizes the affine closure of the cotangent bundle of the minimal nilpotent orbit of $\mathfrak{sl}_n$. This space is a symplectic singularity recently identified by Fu and Liu as a $\mathrm{U}(1)$ hyperk\"ahler quotient of the closure of the minimal nilpotent orbit of $\mathfrak{so}_{2n+2}$. The theory arises by gauging an $\mathrm{SO}(2)\cong\mathrm{U}(1)$ subgroup of the flavour symmetry of $\mathrm{SU}(2)$ SQCD with $n+1$ flavours.

The Hilbert series is computed and the stratification is determined. A non-simply laced magnetic quiver is proposed whose Coulomb branch reproduces the same singularity. Evidence is thereby provided for a mirror pair involving a non-simply laced quiver, further tested through quiver subtraction and Hasse diagram inversion.

A related $\mathbb{Z}_2$ quotient of the magnetic lattice is also analysed, and the exceptional behaviour in the case $n=2$, where $A_1 \cong C_1$, is explained.

This construction provides a concrete example in which the Higgs-branch structure associated with a non-simply laced magnetic quiver can be inferred and validated through its mirror dual.
}

\begin{document}

\maketitle

\section{Introduction}

Moduli spaces of supersymmetric gauge theories in three dimensions provide a powerful bridge between quantum field theory, representation theory, and algebraic geometry. In particular, the Higgs and Coulomb branches of $3\text{d}$ $\mathcal{N}=4$ theories furnish large classes of hyperk\"ahler varieties, many of which arise as symplectic singularities. Quiver gauge theories provide a systematic framework for constructing and studying such moduli spaces, leading to a detailed understanding of nilpotent orbit closures, Slodowy slices, and related varieties, for example \cite{kronheimer1990instantons,Nakajima:1994nid,Gaiotto:2008ak,Cremonesi:2014kwa,Cabrera:2018ldc}.

A central theme in this program is the role of quivers as combinatorial data encoding the geometry of moduli spaces. In the simply laced case, both Higgs and Coulomb branches admit well-understood descriptions. Higgs branches are realized as hyperk\"ahler quotients, while Coulomb branches can be computed using monopole operators \cite{Borokhov:2002cg,Cremonesi:2013lqa}. Mirror symmetry exchanges these two branches and provides a powerful duality relating distinct gauge theories with identical moduli spaces \cite{Intriligator:1996ex}.

% Closely related to mirror symmetry is the mathematical notion of symplectic duality, which provides a unifying framework for understanding the relationship between Higgs and Coulomb branches. Within this perspective, the stratifications of these moduli spaces are encoded by Hasse diagrams, whose nodes correspond to symplectic leaves and whose edges are labeled by transverse slices. In many examples studied in the physics literature, notably in the work of Hanany and collaborators, the Hasse diagrams of Higgs and Coulomb branches are related by an inversion operation: the partial order is reversed, and each slice is replaced by a space of the opposite type (Higgs or Coulomb). When applicable, this inversion allows one to reconstruct the local geometry of one branch from the other. Importantly, the inversion acts not only on the topology of the diagram but also on the gauge-theoretic data labeling the leaves and slices, making it a powerful tool for analyzing moduli spaces even in the absence of a direct Lagrangian description.

Closely related to mirror symmetry is the notion of symplectic duality \cite{Braden:2022qua,Kamnitzer:2022sym,Webster:2023dim}, which provides a unifying framework for understanding the relationship between Higgs and Coulomb branches.
In many cases, the Hasse diagrams of Higgs and Coulomb branches are related by an inversion operation, whereby the partial order is reversed and each slice is replaced by a space of the opposite type \cite{Grimminger:2020dmg}. When applicable, this inversion allows one to reconstruct the local geometry of one branch from the other, and can be viewed as a physical manifestation of symplectic duality.

Recently, it has been shown by Fu and Liu that the affine closure of the cotangent bundle to the minimal nilpotent orbit of $\mathfrak{sl}_n$ defines a symplectic singularity which is isomorphic to a $\mathrm{U}(1)$ hyperk\"ahler quotient of the closure of the minimal nilpotent orbit of $\mathfrak{so}_{2n+2}$ \cite{Fu:2026con}. This construction suggests the existence of an underlying gauge-theoretic realization in terms of $3\text{d}$ $\mathcal{N}=4$ quiver theories.

In this paper, such a realization is presented. A $3$d $\mathcal{N}=4$ theory is identified whose Higgs branch reproduces this symplectic singularity. The corresponding magnetic quiver is proposed as a non-simply laced quiver whose Coulomb branch yields the same space. Magnetic quivers have recently emerged as an effective tool for encoding the geometry of Higgs branches, particularly in situations where a direct Lagrangian description is not available \cite{Cabrera:2018jxt,Hanany:2018vph,Cabrera:2019izd,Cabrera:2019dob}.

The main result of this paper can be summarized as follows. We construct a $3$d $\mathcal{N}=4$ gauge theory $\mathcal{T}$ whose Higgs branch realizes the symplectic singularity
\[
\overline{T^*\mathcal{O}^\mathrm{min}_{\mathfrak{sl}_n}}^{\mathrm{aff}},
\]
which is the affine closure of the cotangent bundle of the minimal nilpotent orbit of $\mathfrak{sl}_n$,
and propose a mirror theory $\mathcal{T}^\vee$ given by a non-simply laced quiver, whose Coulomb branch reproduces the same space. This proposal is supported by a detailed matching of Hilbert series and Hasse diagrams, and further allows for the reconstruction of the Higgs branch of $\mathcal{T}^\vee$ via symplectic duality.

Non-simply laced quivers present a substantial challenge in this framework. In contrast to simply laced quivers, they generally do not admit a perturbative Lagrangian description, and their Higgs branches cannot be constructed using conventional methods. Nevertheless, their Coulomb branches can still be accessed through generalizations of the monopole formula, providing a powerful indirect probe of their geometry \cite{Cremonesi:2013lqa,Cremonesi:2014xha}. A rigorous mathematical definition of Coulomb branches for non-simply laced quivers can be formulated using the BFN construction \cite{Braverman:2016wma} together with later extensions \cite{Nakajima:2020cou}.
Related constructions of non-simply laced quivers arise from folding of Dynkin diagrams and from brane configurations involving orientifolds. The present work differs from these approaches in that it focuses on the reconstruction of Higgs branch geometry via Coulomb branch data and symplectic duality, rather than providing a direct Lagrangian description.

Attempts to extend quiver constructions beyond the simply laced case appear in various contexts.
 In particular, non-simply laced quivers naturally arise from folding procedures of simply laced Dynkin diagrams, as well as from brane constructions involving orientifolds or outer automorphism twists. While these approaches successfully reproduce certain aspects of the corresponding Coulomb branches, a direct Lagrangian description of the associated Higgs branches remains elusive. In abelian cases, non-simply laced edges can be interpreted in terms of hypermultiplets with higher charges, leading to well-defined hypertoric varieties. However, for non-abelian gauge theories such an interpretation generally fails, as rescaling the weights of a representation does not produce a valid representation of the gauge group. Alternative approaches based on generalized quiver techniques, such as quiver subtraction and contraction \cite{Cabrera:2018ann}, as well as the use of highest weight generating functions \cite{Hanany:2014dia,Hanany:2018vph}, provide indirect access to the structure of these moduli spaces. Nevertheless, a systematic framework for constructing Higgs branches associated with non-simply laced quivers remains an open problem.

The example presented here provides a concrete realization of a non-simply laced magnetic quiver whose associated symplectic singularity can be analyzed explicitly. In particular, the stratification of the singular space can be matched with the expected structure arising from the geometric construction. This constitutes a first explicit instance in which a non-simply laced magnetic quiver admits a detailed, stratum-by-stratum verification of its associated Higgs branch geometry.

This work contributes to the broader program of extending quiver-based constructions of moduli spaces beyond the simply laced setting, with the aim of achieving a systematic understanding of non-simply laced symplectic singularities and their physical realizations.

\paragraph{Status of the mirror proposal.}
The mirror theory $\mathcal{T}^\vee$ proposed in this work is not derived from a first-principles construction such as a brane realization or folding procedure. Instead, it is obtained by requiring consistency with the stratification of the Higgs branch of $\mathcal T$ and the expected structure of transverse slices. The proposal is supported by several nontrivial checks, including matching of Hilbert series for low ranks, agreement of Hasse diagrams, and compatibility with symplectic duality. In this sense, the mirror pair $(\mathcal T,\mathcal T^\vee)$ should be regarded as a conjectural pair, supported by strong evidence rather than a derivation from an underlying microscopic construction.

\paragraph{Organisation of the paper.}
The rest of this paper is organised as follows. In Section \ref{sec_2}, the theory $\mathcal{T}$ is introduced as a $\mathrm{U}(1)\cong\mathrm{SO}(2)$ gauging of $\mathrm{SU}(2)$ SQCD with $n+1$ flavours, and its Higgs branch is shown to realize $\overline{T^*\mathcal{O}^{\mathrm{min}}_{\mathfrak{sl}_n}}^{\mathrm{aff}}$. The corresponding Hilbert series is computed, and the Higgsing pattern is analysed, thereby giving the stratification of the singularity. In Section \ref{sec_3}, the non-simply laced mirror theory $\mathcal{T}^\vee$ is proposed. The abelian mirror of the theory $\mathcal{A}$ is first reviewed, after which the Coulomb branch of $\mathcal{T}^\vee$ is analysed using the monopole formula. Hasse-diagram inversion, motivated by symplectic duality, is then used to reconstruct the expected Higgs branch of $\mathcal{T}^\vee$, which is compared with the Coulomb branch of $\mathcal{T}$. In Section \ref{sec_4}, a related discrete quotient obtained from alternative ungauging choices is studied, and its interpretation as a $\mathbb{Z}_2$ quotient on the Coulomb branch is explained. In Section \ref{sec_5}, the exceptional case $n=2$ is treated, where $A_1\cong C_1$ and the higher-rank construction no longer applies directly.

\section{Gauging from $D$ to $A$}
\label{sec_2}

The theory $\mathcal{T}$ is defined as follows:
\begin{equation}
    \mathcal{T}=\raisebox{-0.5\height}{\begin{tikzpicture}
    \node[gauge,label=below:{$\mathrm{SU}(2)$}] (0) at (0,0) {};
    \node[gauge,label=below:{$\mathrm{SO}(2)$}] (1) at (1.5,0) {};
    \node[flavour,label=above:{$\mathrm{O}(2)$}] (0a) at (0,1) {};
    \draw (0a)--(0);
    \node at (0.75,0.3) {$n$};
    %\draw (0a) .. controls (4,1) .. (4);
    \draw[transform canvas={yshift=2pt}] (0)--(1);
    \draw[transform canvas={yshift=0pt}] (0)--(1);
    \draw[transform canvas={yshift=-2pt}] (0)--(1);
    %\draw [thick,decoration={brace,mirror,raise=0.8cm},decorate] (1) -- (4) node [pos=0.5,anchor=north,yshift=-0.85cm] {$n-2$}; 
    %\draw (-.2,.6)--(0,0.4)--(.2,.6);
     %\node at (3.5,.5) {$k$};
     \end{tikzpicture}},
\label{eq_T}
\end{equation}
where $n\geq3$. The continuous flavour symmetry is $\mathrm {SO}(2)\times \mathrm{SU}(n)$.
The theory $\mathcal{T}$ is a $\mathrm{U}(1)\cong\mathrm{SO}(2)$ gauging of $\mathrm{SU}(2)$ with $n+1$ flavours. The embedding of this $\mathrm{U}(1)$ into $\mathrm{SO}(2n+2)$ is determined uniquely by its normaliser which is the flavour symmetry of $\mathcal{T}$.
On the Higgs branch, the $\mathrm{U}(1)$ gauging is exactly a $\mathbb{C}^\times$ Hyperk\"ahler quotient. The Higgs branch of $\mathrm{SU}(2)$ with $n+1$ flavours is well known to be $d_{n+1}:=\overline{\mathcal{O}^\mathrm{min}_{\mathfrak{so}_{2n+2}}}$. Together, the following relation holds:
\begin{equation}
     \mathcal{H}(\mathcal{T})\cong d_{n+1}///\mathbb{C}^\times.
\label{eq_idHdc}
\end{equation}
This $\mathbb{C}^\times$ action agrees with the action specified in \cite{Fu:2026con}, hence

%One can check this $\mathbb{C}^\times$ action on $d_{n+1}$ induced by the $\mathrm{U}(1)$ agrees with the $\mathbb{C}^\times$ action specified in \cite{Fu:2026con}, hence,
\begin{equation}
     \mathcal{H}(\mathcal{T})\cong\overline{T^*\mathcal{O}^\mathrm{min}_{\mathfrak{sl}_n}}^{\mathrm{aff}}.
\label{eq_HTaffine}
\end{equation}
Equation \eqref{eq_HTaffine} constitutes the central identification of this work, relating the Higgs branch of the gauge theory $\mathcal{T}$ to the symplectic singularity constructed in \cite{Fu:2026con}.

\subsection{Hilbert series}
An important characteristic for a symplectic singularity is its Hilbert series (HS), which is counting the degrees of invariant polynomials. The HS of Higgs branch counts the gauge invariant operators (GIO) and can be computed through the Molien-Weyl integral\footnote{Plethystic exponential (PE) for a function $f$ which is vanishing at the origin $f(0)=0$, is defined as $\mathrm{PE}[f(x)]=\mathrm{exp}(\sum_{k=1}^{\infty}\frac{f(x^k)}{k})$. In particular, $\mathrm{PE}[a t^n]=(1-t^n)^{-a}$.}:
\begin{equation}
    \oint d\mu_G\ \mathrm{PE}[\chi_{hyper}t-\chi_{adj}t^2],
\end{equation}
where $d\mu_G$ is the Haar measure of gauge group $G$, $\chi_{hyper}t$ is the contribution from hypermultiplets, and $-\chi_{adj}t^2$ is the contribution from $F$-term equations. In particular, for $\mathcal{H}(\mathcal{T})$:
\begin{equation}
    \mathrm{HS}_{\mathcal{H}}(n)=\oint  \frac{1-x^2}{x}\frac{1}{q}dx\ dq\ \mathrm{PE}[n(x+\frac{1}{x})(q+\frac{1}{q})t+2(x+\frac{1}{x})t-(x^2+1+\frac{1}{x^2})t^2-t^2].
\label{eq_HSHT}
\end{equation}
For $n=3,4$, the results are:
\begin{equation}
    \mathrm{HS}_{\mathcal{H}}(3)=
\frac{1+5t^2+26t^4+55t^6+78t^8+55t^{10}+26t^{12}+5t^{14}+t^{16}}
{(1-t^2)^4(1-t^4)^4}.
\label{eq_HS3}
\end{equation}
\begin{align}
    \mathrm{HS}_{\mathcal{H}}(4)=
    \mathrm{PE}[6t^2+6t^4]&
    (1+10t^2+78t^4+332t^6+911t^8+1606t^{10}+1956t^{12}
\notag\\&+1606t^{14}+911t^{16}+332t^{18}+78t^{20}+10t^{22}+t^{24})
.
\label{eq_HS4}
\end{align}
The coefficient of the $t^2$ term agrees with the dimension of the flavour symmetry.

%It is easy to check that the coefficient of the \(t^2\) term agrees with the dimension of the flavour symmetry.

\subsection{Higgs mechanism}

The Higgsing procedure considered in this work proceeds by moving along suitably chosen directions in the Higgs branch of the theory $\mathcal{T}$, such that the gauge symmetry is partially or completely broken while preserving supersymmetry. At each step, vacuum expectation values are assigned to hypermultiplet scalars in a way that minimizes the breaking pattern, leading to a controlled descent through a sequence of gauge theories. This process can be organized in terms of transverse slices, which capture the local structure of the Higgs branch near a given vacuum and are themselves described by lower-rank gauge theories. Crucially, the Higgsing procedure captures the stratification of symplectic leaves of underlying symplectic singularity \cite{Bourget:2019aer}.

In particular, the theory $\mathcal{T}$ can be Higgsed to an abelian theory $\mathcal{A}$ in a minimal way. The resulting theory $\mathcal{A}$ has gauge group $\mathrm{U}(1)$ and $n$ hypermultiplets with charge matrix
\begin{equation}
    \mathbf{q}=\begin{pmatrix}
        1,1,2,\dots,2
    \end{pmatrix}^\mathsf{T}.
\end{equation}
This theory also admits a quiver,
\begin{equation}
    \mathcal{A}=\raisebox{-0.5\height}{\begin{tikzpicture}
    \node[gauge,label=below:{$1$}] (0) at (0,0) {};
    \node[flavour,label=below:{$n\!-\!2$}] (1) at (1,0) {};
    \node[flavour,label=above:{$2$}] (0a) at (0,1) {};
    \draw (0a)--(0)  (0.4,0.2)--(0.6,0)--(0.4,-0.2);
    \draw[transform canvas={yshift=1.5pt}] (0)--(1);
    \draw[transform canvas={yshift=-1.5pt}] (0)--(1);
     \end{tikzpicture}}.
\label{eq_A}
\end{equation}

The Higgs branch of $\mathcal{A}$ is a hypertoric variety $\bar h_{\mathbf{q}^\mathsf{T}}$ as described in \cite{Grimminger:2025fgj}. For $n=3$, this space is an isolated singularity. For $n\geq4$, it admits a further Higgsing to a discrete gauge theory $\mathcal{D}$ with gauge group $\mathbb{Z}_2$ and two hypermultiplets,
\begin{equation}
    \mathcal{D}=\raisebox{-0.5\height}{\begin{tikzpicture}
    \node[gauge,label=below:{$\mathbb{Z}_2$}] (0) at (0,0) {};
    \node[flavour,label=below:{$2$}] (1) at (1,0) {};
    \draw (0)--(1);
     \end{tikzpicture}}.
\label{eq_D}
\end{equation}
The Higgs branch of $\mathcal{D}$ is given by
\[
c_2 := \mathbb{C}^4/\mathbb{Z}_2.
\]

The local structure of the Higgs branch is encoded in transverse slices. The transverse slice from $\mathcal{T}$ to $\mathcal{A}$ is described by the Higgs branch of the theory $\mathcal{N}_{\mathcal{T}}(\mathcal{A})$, which is a $\mathrm{U}(1)$ gauge theory with $n$ hypermultiplets,
\begin{equation}
    \mathcal{N}_{\mathcal{T}}(\mathcal{A})=\raisebox{-0.5\height}{\begin{tikzpicture}
    \node[gauge,label=below:{$1$}] (0) at (0,0) {};
    \node[flavour,label=below:{$n$}] (1) at (1,0) {};
    \draw (0)--(1);
     \end{tikzpicture}},
\end{equation}
whose Higgs branch is
\[
a_{n-1}:=\overline{\mathcal{O}^\mathrm{min}_{\mathfrak{sl}_{n}}}.
\]

Similarly, the transverse slice from $\mathcal{A}$ to $\mathcal{D}$ is described by the Higgs branch of
\begin{equation}
    \mathcal{N}_{\mathcal{A}}(\mathcal{D})=\raisebox{-0.5\height}{\begin{tikzpicture}
    \node[gauge,label=below:{$1$}] (0) at (0,0) {};
    \node[flavour,label=below:{$n\!-\!2$}] (1) at (1,0) {};
    \draw (0)--(1);
     \end{tikzpicture}},
\label{eq_NAD}
\end{equation}
whose Higgs branch is $a_{n-3}$.

The transverse slice from $\mathcal{T}$ directly to $\mathcal{D}$ is described by
\begin{equation}
    \mathcal{N}_{\mathcal{T}}(\mathcal{D})=\raisebox{-0.5\height}{\begin{tikzpicture}
    \node[gauge,label=below:{$2$}] (0) at (0,0) {};
    \node[flavour,label=below:{$n$}] (1) at (1,0) {};
    \draw (0)--(1);
     \end{tikzpicture}},
\end{equation}
whose Higgs branch is the closure of the next-to-minimal nilpotent orbit,
\[
\overline{\mathcal{O}^\mathrm{n.min}_{\mathfrak{sl}_{n}}}.
\]

The resulting Higgsing pattern can be summarized by the following Hasse diagrams:
\begin{equation}
%\vcenter{\hbox{\begin{tikzpicture}
%        \node[hasse,label=right:{$\mathcal H(\mathcal{T})$}] (0) at (0,0) {};
%        \node[hasse,label=right:{$\varnothing$}] (1) at (0,1.5) {};
%        \draw (0)--(1);
%        \node at (-0.5,0.75) {$c_2$};
%        \node at (0,-1) {$n=2$};
%    \end{tikzpicture}}}
%    \hspace{1.2cm}
        \vcenter{\hbox{\begin{tikzpicture}
        \node[hasse,label=right:{$\mathcal H(\mathcal{T})$}] (0) at (0,0) {};
        \node[hasse,label=right:{$\mathcal H(\mathcal{A})$}] (1) at (0,1.5) {};
        \node[hasse,label=right:{$\varnothing$}] (2) at (0,3) {};
        \draw (0)--(1) (1)--(2);
        \node at (-0.7,2.25) {$\bar h_{(1,1,2)^\mathsf{T}}$};
        \node at (-0.5,0.75) {$a_2$};
        \node at (0,-1) {$n=3$};
    \end{tikzpicture}}}
    \hspace{2cm}
    \vcenter{\hbox{\begin{tikzpicture}
        \node[hasse,label=right:{$\mathcal H(\mathcal{T})$}] (0) at (0,0) {};
        \node[hasse,label=right:{$\mathcal H(\mathcal{A})$}] (1) at (0,1.5) {};
        \node[hasse,label=right:{$\mathcal H(\mathcal{D})$}] (2) at (0,3) {};
        \node[hasse,label=right:{$\varnothing$}] (3) at (0,4.5) {};
        \draw (0)--(1) (1)--(2) (2)--(3);
        \node at (-0.5,3.75) {$c_2$};
        \node at (-0.5,2.25) {$a_{n-3}$};
        \node at (-0.5,0.75) {$a_{n-1}$};
        \node at (0,-1) {$n\geq4$};
        \draw [thick,decoration={brace,mirror,raise=1.5cm},decorate] (1) -- (3)
        node [pos=0.5,anchor=north,xshift=2.7cm,yshift=0.2cm]
        {$\bar h_{(1,1,2,\dots,2)^\mathsf{T}}$};
        \draw [thick,decoration={brace,mirror,raise=1cm},decorate] (2) -- (0)
        node [pos=0.5,anchor=north,xshift=-2cm,yshift=0.2cm]
        {$\overline{\mathcal{O}^\mathrm{n.min}_{\mathfrak{sl}_{n}}}$};
    \end{tikzpicture}}}\qquad,
    \label{eq_hasseT}
\end{equation}
where each node labels a symplectic leaf and the corresponding transverse slice to the top. The Higgsing proceeds from the bottom, corresponding to the most singular point, to the top, which is smooth. 
This Higgsing structure is in agreement with the stratification into symplectic leaves of $\overline{T^*\mathcal{O}^\mathrm{min}_{\mathfrak{sl}_n}}^{\mathrm{aff}}$ \cite{Fu:2026con}.

\section{Non-simply laced $\mathcal{T}^\vee$ and mirror symmetry}
\label{sec_3}

This section discusses a non-simply laced theory $\mathcal{T}^\vee$, which is proposed to be the 3d mirror of $\mathcal{T}$. The Coulomb branch $\mathcal{C}(\mathcal{T}^\vee)$ can be analysed from monopole formula while the Higgs branch $\mathcal{T}^\vee$ has no perturbatively description.

To describe $\mathcal{H}(\mathcal{T}^\vee)$, a central role will be played by the following set of building-block theories:
\begin{equation}
    \mathcal{G}^\vee:=\{\mathcal{D}^\vee,\ \mathcal{A}^\vee,\ \mathcal{N}_{\mathcal{A}}(\mathcal{D})^\vee,\ \mathcal{N}_{\mathcal{T}}(\mathcal{D})^\vee,\ \mathcal{N}_{\mathcal{T}}(\mathcal{A})^\vee\}.
\end{equation}
These theories encode the local geometry of $\mathcal{C}(\mathcal{T}^\vee)$ and form the building blocks for the inversion procedure.
\subsection{Abelian mirror}
% Before determine the 3d mirror of $\mathcal{T}$, it is useful to firstly study the mirror of $\mathcal{A}$.
% The 3d mirror of an abelian theory can be determined using the methods in \cite{deBoer:1996ck,deBoer:1996mp,Grimminger:2025fgj}.
% The mirror theory $\mathcal{A}^\vee$ of theory $\mathcal{A}$ has gauge group $\mathrm{U}(1)^{n-1}$ and $n$ hypers, whose charge matrix is given by an $(n-1)\times n$ matrix:
Before determining the 3d mirror of \(\mathcal{T}\), it is useful to first study the mirror of \(\mathcal{A}\).
The 3d mirror of an abelian theory can be determined using the methods in \cite{deBoer:1996ck,deBoer:1996mp,Grimminger:2025fgj}.
The mirror theory \(\mathcal{A}^\vee\) of theory \(\mathcal{A}\) has gauge group \(\mathrm{U}(1)^{n-1}\) and \(n\) hypermultiplets, whose charge matrix is given by an \((n-1)\times n\) matrix:
\begin{equation}
    \mathbf{q}^\vee=\begin{pmatrix}
        1&&&&&\\
        -1&2&&&&\\
        &-1&1&&&\\
        &&&\ddots&&\\
        &&&&-1&1\\
        &&&&&-1
    \end{pmatrix}.
\end{equation}
% A free $\mathrm{U}(1)$ can be always be coupled without change the rank of the charge matrix\footnote{It can be treated as a generalisation to Crawley-Boevey move defined in math literature \cite{Crawley-Boevey1}. See also \cite{Grimminger:2025fgj}.}, the resulting theory has the same Higgs branch and the same Coulomb branch up to a factor $T^*\mathbb{C}^\times$. Since the singularity remains the same, the two theories before or after adding the free $\mathrm{U}(1)$ are not distinguished for the purpose of this paper.
A free $\mathrm{U}(1)$ factor can always be coupled without changing the rank of the charge matrix.\footnote{This can be viewed as a generalization of the Crawley--Boevey move in the mathematical literature \cite{Crawley-Boevey1}. See also \cite{Grimminger:2025fgj}.} The resulting theory has the same Higgs branch and Coulomb branch, up to an additional factor of $T^*\mathbb{C}^\times$. Since the underlying singularity remains unchanged, the theories before and after adding the free $\mathrm{U}(1)$ are not distinguished for the purposes of this paper.
After adding the free \(\mathrm{U}(1)\), the theory \(\mathcal{A}\) also admits a quiver, but it is unframed.
\begin{equation}
    \mathcal{A}^\vee=
    \raisebox{-0.5\height}{\begin{tikzpicture}
    \node[gauge,label=below:{1}] (0) at (0,0) {};
    \node[gauge,label=below:{1}] (1) at (1,0) {};
    \node (2) at (2,0) {$\cdots$};
    \node[gauge,label=below:{1}] (3) at (3,0) {};
    \node[gauge,label=below:{1}] (4) at (4,0) {};
    %\node[gauge,label=below:{1}] (5) at (5,0) {};
    %\node[flavour,label=above:{1}] (4a) at (4,1) {};
    %\node[flavour,label=above:{1}] (3a) at (3,1) {};
    \node[gauge,label=above:{1}] (2a) at (2,1) {};
    %\node[flavour,label=above:{1}] (1a) at (1,1) {};
    %\node[gauge,label=above:{1}] (0a) at (0,1) {};
    \draw (0)--(1)--(2)--(3)--(4);
    \draw[double,double distance=3pt] (0)--(2a)--(4);
    \draw[shift={(1.1,0.55)},rotate=26.565] (-0.212,0.212)--(0,0)--(-0.212,-0.212);
    \draw[shift={(2.9,0.55)},rotate=153.435](-0.212,0.212)--(0,0)--(-0.212,-0.212);
    %\draw (0a) .. controls (4,1) .. (4);
    \draw [thick,decoration={brace,mirror,raise=0.8cm},decorate] (0) -- (4) 
node [pos=0.5,anchor=north,yshift=-0.85cm] {$n-1$}; 
    %\draw (-.2,.6)--(0,0.4)--(.2,.6);
     %\node at (3.5,.5) {$k$};
     \end{tikzpicture}}.
     \label{Achech}
\end{equation}

%\begin{tikzpicture}

% Bottom nodes
%\node[gauge,label=below:{1}] (0) at (0,0) {};
%\node[gauge,label=below:{1}] (1) at (1,0) {};
%\node (dots) at (2,0) {$\cdots$};
%\node[gauge,label=below:{1}] (n2) at (3,0) {};
%\node[gauge,label=below:{1}] (n1) at (4,0) {};

% Top node
%\node[gauge,label=above:{1}] (top) at (2,1.5) {};

% Bottom chain (n-1 nodes)
%\draw (0)--(1)--(dots)--(n2)--(n1);

% Non-simply laced edges (double with arrows pointing up)
%\draw[double,double distance=1.2pt,->] (0) -- (top);
%\draw[double,double distance=1.2pt,->] (n1) -- (top);

% Brace showing n-1 nodes
%\draw[thick,decoration={brace,mirror,raise=0.8cm},decorate]
%(0) -- (n1)
%node[pos=0.5,anchor=north,yshift=-0.85cm] {$n-1$};

%\end{tikzpicture}

Similarly, the mirror theory $\mathcal{D}^\vee$ of theory $\mathcal{D}$ can be computed via the methods in \cite{Grimminger:2025fgj,Bourget:2025wsp}. Being a minimal nilpotent orbit, it admits a unframed quiver as an affine Dynkin diagram
\begin{equation}
    \mathcal{D}^\vee=
    \raisebox{-0.5\height}{\begin{tikzpicture}
    \node[gauge,label=below:{1}] (0) at (0,0) {};
    \node[gauge,label=below:{1}] (1) at (1,0) {};
    \node[gauge,label=below:{1}] (2) at (2,0) {};
    \draw (0.4,0.2)--(0.6,0)--(0.4,-0.2) (1.6,0.2)--(1.4,0)--(1.6,-0.2);
    %\draw (0a) .. controls (4,1) .. (4);
    \draw[transform canvas={yshift=-1.5pt}] (0)--(1)--(2);
    \draw[transform canvas={yshift=1.5pt}] (0)--(1)--(2);
    %\draw (-.2,.6)--(0,0.4)--(.2,.6);
     %\node at (3.5,.5) {$k$};
     \end{tikzpicture}}.
     \label{affinec2}
\end{equation}
After quiver subtraction of $\mathcal{D}^\vee$ from $\mathcal{A}^\vee$, the resulting quiver is precisely the mirror of $\mathcal{N}_{\mathcal{A}}(\mathcal{D})$,
\begin{equation}
    \mathcal{A}^\vee-\mathcal{D}^\vee=
    \raisebox{-0.5\height}{\begin{tikzpicture}
    %\node[gauge,label=below:{1}] (0) at (0,0) {};
    \node[gauge,label=below:{1}] (1) at (1,0) {};
    \node (2) at (2,0) {$\cdots$};
    \node[gauge,label=below:{1}] (3) at (3,0) {};
    %\node[gauge,label=below:{2}] (4) at (4,0) {};
    %\node[gauge,label=below:{1}] (5) at (5,0) {};
    %\node[flavour,label=above:{1}] (4a) at (4,1) {};
    %\node[flavour,label=above:{1}] (3a) at (3,1) {};
    \node[gauge,label=above:{1}] (2a) at (2,1) {};
    %\node[flavour,label=above:{1}] (1a) at (1,1) {};
    %\node[gauge,label=above:{1}] (0a) at (0,1) {};
    \draw (2a)--(1)--(2)--(3)--(2a);
    %\draw (0a) .. controls (4,1) .. (4);
    \draw [thick,decoration={brace,mirror,raise=0.8cm},decorate] (1) -- (3) 
node [pos=0.5,anchor=north,yshift=-0.85cm] {$n-3$}; 
    %\draw (-.2,.6)--(0,0.4)--(.2,.6);
     %\node at (3.5,.5) {$k$};
     \end{tikzpicture}}=\mathcal{N}_{\mathcal{A}}(\mathcal{D})^\vee.
     \label{affinean3}
\end{equation}
\paragraph{Quiver contraction}
As quiver \eqref{affinec2} is a subquiver of quiver \eqref{Achech} one gets quiver \eqref{affinean3} by contracting \eqref{affinec2} into a single node. This procedure is equivalent to the notion of quiver subtraction, but will henceforth be called quiver contraction.
% Different from Higgsing, the quiver contraction on Coulomb branch goes from top to bottom. In the other word, after quiver subtraction, one reach a lower leaf whose closure is described by the Coulomb branch of the remaining quiver. The transverse slice is described by the Coulomb branch of the contracted quiver.
Unlike Higgsing, quiver contraction on the Coulomb branch proceeds from top to bottom. In other words, after quiver subtraction, one reaches a lower leaf whose closure is described by the Coulomb branch of the remaining quiver. The transverse slice is described by the Coulomb branch of the contracted quiver.

Due to the mirror relation, $\mathcal{H}(\mathcal{A})$ and $\mathcal{C}(\mathcal{A}^\vee)$ are equal, so are their Hasse diagrams:
\begin{equation}
        \vcenter{\hbox{\begin{tikzpicture}
        \node[hasse,label=right:{$\mathcal H(\mathcal{A})$}] (0) at (0,0) {};
        \node[hasse,label=right:{$\mathcal H(\mathcal{D})$}] (1) at (0,1.5) {};
        \node[hasse,label=right:{$\varnothing$}] (2) at (0,3) {};
        \draw (0)--(1) (1)--(2);
        \node at (-0.5,2.25) {$c_2$};
        \node at (-0.5,0.75) {$a_{n-1}$};
        \node at (0,3.5) {};
        %\node at (0,-1) {$n=3$};
        \draw [thick,decoration={brace,mirror,raise=1cm},decorate] (1) -- (0) node [pos=0.5,anchor=north,xshift=-2cm,yshift=0.2cm] {$\mathcal H(\mathcal{N}_{\mathcal{A}}(\mathcal{D}))$};
    \end{tikzpicture}}}
    \hspace{1cm} = \hspace{1cm}
    \vcenter{\hbox{\begin{tikzpicture}
        \node[hasse,label=right:{$\varnothing$}] (0) at (0,0) {};
        \node[hasse,label=right:{$\mathcal C(\mathcal{N}_{\mathcal{A}}(\mathcal{D})^\vee)$}] (1) at (0,1.5) {};
        \node[hasse,label=right:{$\mathcal C(\mathcal{A}^\vee)$}] (2) at (0,3) {};
        \draw (0)--(1) (1)--(2);
        \node at (-0.5,2.25) {$c_2$};
        \node at (-0.5,0.75) {$a_{n-1}$};
        \node at (0,3.5) {};
        %\node at (0,-1) {$n\geq4$};
        \draw [thick,decoration={brace,mirror,raise=0.8cm},decorate] (2) -- (1) node [pos=0.5,anchor=north,xshift=-1.8cm,yshift=0.2cm] {$\mathcal C(\mathcal{D}^\vee)$}; 
    \end{tikzpicture}}}\qquad.
\end{equation}
By 3d mirror symmetry for abelian theories $\mathcal{C}(\mathcal{A})$ and $\mathcal{H}(\mathcal{A}^\vee)$ are also equal, and are discussed in section \ref{sec_symplecticdual}.

%For abelian theories, a non-simply laced edge is just a hypermultiplet whose weights under certain $\mathrm{U}(1)$s are larger than $1$, the theories is perfectly well-defined. However a non-simply laced edge from to a non-abelian gauge group is not well-defined, the weight of a representation of a semi-simple lie group scaled by an arbitrary integer is no longer a representation. For example, there is only one non-trivial 2 dimensional representation of $\mathrm{SU(2)}$, which is the irrep with weight $\{-1,1\}$. A 2-dim vector space with weight $\{n_1,n_2\}$ is not a representation but a subspace of another representation. 

\subsection{Non-simply laced mirror and Coulomb branch of $\mathcal{T}^\vee$}
\label{sec_nslm}
For non-abelian theories, such as \(\mathcal{T}\), there is no general method to obtain their mirrors, for example \(\mathcal{T}^\vee\). However, with the knowledge of the Hasse diagram, an educated guess can be made:
\begin{equation}
    \mathcal{T}^\vee=
    \raisebox{-0.5\height}{\begin{tikzpicture}
    \node[gauge,label=below:{2}] (0) at (0,0) {};
    \node[gauge,label=below:{2}] (1) at (1,0) {};
    \node (2) at (2,0) {$\cdots$};
    \node[gauge,label=below:{2}] (3) at (3,0) {};
    \node[gauge,label=below:{2}] (4) at (4,0) {};
    %\node[gauge,label=below:{1}] (5) at (5,0) {};
    %\node[flavour,label=above:{1}] (4a) at (4,1) {};
    %\node[flavour,label=above:{1}] (3a) at (3,1) {};
    \node[gauge,label=above:{1}] (2a) at (2,1) {};
    %\node[flavour,label=above:{1}] (1a) at (1,1) {};
    %\node[gauge,label=above:{1}] (0a) at (0,1) {};
    \draw (0)--(1)--(2)--(3)--(4);
    \draw[double,double distance=3pt] (0)--(2a)--(4);
    %\draw (0.8,1.1)--(1.1,1.1)--(1.1,0.8) (3.2,1.1)--(2.9,1.1)--(2.9,0.8);
    \draw[shift={(1.1,0.55)},rotate=26.565] (-0.212,0.212)--(0,0)--(-0.212,-0.212);
    \draw[shift={(2.9,0.55)},rotate=153.435](-0.212,0.212)--(0,0)--(-0.212,-0.212);
    %\draw (0a) .. controls (4,1) .. (4);
    \draw [thick,decoration={brace,mirror,raise=0.8cm},decorate] (0) -- (4) 
node [pos=0.5,anchor=north,yshift=-0.85cm] {$n-1$}; 
    %\draw (-.2,.6)--(0,0.4)--(.2,.6);
     %\node at (3.5,.5) {$k$};
     \end{tikzpicture}}.
     \label{eq_Tm}
\end{equation}
The quiver in \eqref{eq_Tm} is proposed based on consistency with the Higgs branch Hasse diagram and the expected structure of transverse slices. Its validity will be justified a posteriori through matching of Hilbert series and Coulomb branch stratification.

The special feature of \(\mathcal{T}^\vee\) is that it is a non-abelian theory with non-simply laced edges.

The Coulomb branch Hasse diagram of $\mathcal{T}^\vee$ can be computed through quiver contraction.
In this case the $\mathcal{D}^\vee$ quiver is the minimal subquiver which can be contracted into a single node of $\mathrm{U}(1)$, the quiver after contraction is as follows:
\begin{equation}
    \mathcal{T}^\vee-\mathcal{D}^\vee=
    \raisebox{-0.5\height}{\begin{tikzpicture}
    \node[gauge,label=below:{1}] (0) at (0,0) {};
    \node[gauge,label=below:{2}] (1) at (1,0) {};
    \node (2) at (2,0) {$\cdots$};
    \node[gauge,label=below:{2}] (3) at (3,0) {};
    \node[gauge,label=below:{1}] (4) at (4,0) {};
    %\node[gauge,label=below:{1}] (5) at (5,0) {};
    %\node[flavour,label=above:{1}] (4a) at (4,1) {};
    %\node[flavour,label=above:{1}] (3a) at (3,1) {};
    \node[gauge,label=above:{1}] (2a) at (2,1) {};
    %\node[flavour,label=above:{1}] (1a) at (1,1) {};
    %\node[gauge,label=above:{1}] (0a) at (0,1) {};
    \draw (0)--(1)--(2)--(3)--(4) (1)--(2a) (3)--(2a);
    %\draw[double,double distance=3pt] (0)--(2a)--(4);
    %\draw (0.8,1.1)--(1.1,1.1)--(1.1,0.8) (3.2,1.1)--(2.9,1.1)--(2.9,0.8);
    %\draw (0a) .. controls (4,1) .. (4);
    \draw [thick,decoration={brace,mirror,raise=0.8cm},decorate] (1) -- (3) node [pos=0.5,anchor=north,yshift=-0.85cm] {$n-3$}; 
    %\draw (-.2,.6)--(0,0.4)--(.2,.6);
     %\node at (3.5,.5) {$k$};
     \end{tikzpicture}}=\mathcal{N}_{\mathcal{T}}(\mathcal{D})^\vee.
\end{equation}
This quiver is precisely the mirror theory of $\mathcal{N}_{\mathcal{T}}(\mathcal{D})$. The transverse slice is given by $\mathcal{C}(\mathcal{D}^\vee)\cong\mathcal{H}(\mathcal{D})$ and the leaf closure is given by $\mathcal{C}(\mathcal{N}_{\mathcal{T}}(\mathcal{D})^\vee)\cong\mathcal{H}(\mathcal{N}_{\mathcal{T}}(\mathcal{D}))$.

From $\mathcal{N}_{\mathcal{T}}(\mathcal{D})^\vee$, the minimal subquiver to contract is $\mathcal{N}_{\mathcal{A}}(\mathcal{D})^\vee$. After contraction,
\begin{equation}
    \mathcal{N}_{\mathcal{T}}(\mathcal{D})^\vee-\mathcal{N}_{\mathcal{A}}(\mathcal{D})^\vee=
    \raisebox{-0.5\height}{\begin{tikzpicture}
    \node[gauge,label=below:{1}] (0) at (0,0) {};
    \node[gauge,label=below:{1}] (1) at (1,0) {};
    \node (2) at (2,0) {$\cdots$};
    \node[gauge,label=below:{1}] (3) at (3,0) {};
    \node[gauge,label=below:{1}] (4) at (4,0) {};
    %\node[gauge,label=below:{1}] (5) at (5,0) {};
    %\node[flavour,label=above:{1}] (4a) at (4,1) {};
    %\node[flavour,label=above:{1}] (3a) at (3,1) {};
    \node[gauge,label=above:{1}] (2a) at (2,1) {};
    %\node[flavour,label=above:{1}] (1a) at (1,1) {};
    %\node[gauge,label=above:{1}] (0a) at (0,1) {};
    \draw (2a)--(0)--(1)--(2)--(3)--(4)--(2a);
    %\draw (0a) .. controls (4,1) .. (4);
    \draw [thick,decoration={brace,mirror,raise=0.8cm},decorate] (0) -- (4) 
node [pos=0.5,anchor=north,yshift=-0.85cm] {$n-1$}; 
    %\draw (-.2,.6)--(0,0.4)--(.2,.6);
     %\node at (3.5,.5) {$k$};
     \end{tikzpicture}}=\mathcal{N}_{\mathcal{T}}(\mathcal{A})^\vee.
\end{equation}
This quiver is precisely the mirror theory of $\mathcal{N}_{\mathcal{T}}(\mathcal{A})$.

Non-minimally, the \(\mathcal{A}^\vee\) subquiver can be directly contracted into a single \(\mathrm{U}(1)\) node. The quiver after contraction is as follows:
\begin{equation}
    \mathcal{T}^\vee-\mathcal{A}^\vee=
    \raisebox{-0.5\height}{\begin{tikzpicture}
    \node[gauge,label=below:{1}] (0) at (0,0) {};
    \node[gauge,label=below:{1}] (1) at (1,0) {};
    \node (2) at (2,0) {$\cdots$};
    \node[gauge,label=below:{1}] (3) at (3,0) {};
    \node[gauge,label=below:{1}] (4) at (4,0) {};
    %\node[gauge,label=below:{1}] (5) at (5,0) {};
    %\node[flavour,label=above:{1}] (4a) at (4,1) {};
    %\node[flavour,label=above:{1}] (3a) at (3,1) {};
    \node[gauge,label=above:{1}] (2a) at (2,1) {};
    %\node[flavour,label=above:{1}] (1a) at (1,1) {};
    %\node[gauge,label=above:{1}] (0a) at (0,1) {};
    \draw (2a)--(0)--(1)--(2)--(3)--(4)--(2a);
    %\draw (0a) .. controls (4,1) .. (4);
    \draw [thick,decoration={brace,mirror,raise=0.8cm},decorate] (0) -- (4) 
node [pos=0.5,anchor=north,yshift=-0.85cm] {$n-1$}; 
    %\draw (-.2,.6)--(0,0.4)--(.2,.6);
     %\node at (3.5,.5) {$k$};
     \end{tikzpicture}}=\mathcal{N}_{\mathcal{T}}(\mathcal{A})^\vee.
\end{equation}
The transverse slice is given by $\mathcal{C}(\mathcal{A}^\vee)\cong\mathcal{H}(\mathcal{A})$ and the leaf closure is given by $\mathcal{C}(\mathcal{N}_{\mathcal{T}}(\mathcal{A})^\vee)\cong\mathcal{H}(\mathcal{N}_{\mathcal{T}}(\mathcal{A}))$.

The Hasse diagram of $\mathcal{C}(\mathcal{T}^\vee)$ agrees with the Higgs branch Hasse diagram of $\mathcal{H}(\mathcal{T})$:
\begin{equation}
        \vcenter{\hbox{\begin{tikzpicture}
        \node[hasse,label=right:{$\mathcal H(\mathcal{T})$}] (0) at (0,0) {};
        \node[hasse,label=right:{$\mathcal H(\mathcal{A})$}] (1) at (0,1.5) {};
        \node[hasse,label=right:{$\mathcal H(\mathcal{D})$}] (2) at (0,3) {};
        \node[hasse,label=right:{$\varnothing$}] (3) at (0,4.5) {};
        \draw (0)--(1) (1)--(2) (2)--(3);
        \node at (-0.5,3.75) {$c_2$};
        \node at (-0.5,2.25) {$a_{n-3}$};
        \node at (-0.5,0.75) {$a_{n-1}$};
        \node at (0,3.5) {};
        %\node at (0,-1) {$n=3$};
        %\draw [thick,decoration={brace,mirror,raise=.5cm},decorate] (3) -- (1) node [pos=0.5,anchor=north,xshift=-1.5cm,yshift=0.2cm] {$\mathcal H(\mathcal{N}_{\mathcal{A}}(\mathcal{D}))$};
        \draw [thick,decoration={brace,mirror,raise=1.5cm},decorate] (0) -- (2) node [pos=0.5,anchor=north,xshift=2.7cm,yshift=0.2cm] {$\mathcal H(\mathcal{N}_{\mathcal{T}}(\mathcal{D}))$};
        \draw [thick,decoration={brace,mirror,raise=1cm},decorate] (2) -- (1) node [pos=0.5,anchor=north,xshift=-2cm,yshift=0.2cm] {$\mathcal H(\mathcal{N}_{\mathcal{A}}(\mathcal{D}))$};
        \draw [thick,decoration={brace,mirror,raise=1cm},decorate] (1) -- (0) node [pos=0.5,anchor=north,xshift=-2cm,yshift=0.2cm] {$\mathcal H(\mathcal{N}_{\mathcal{T}}(\mathcal{A}))$};
    \end{tikzpicture}}}
    \hspace{0.4cm} = \hspace{0.4cm}
    \vcenter{\hbox{\begin{tikzpicture}
        \node[hasse,label=right:{$\varnothing$}] (0) at (0,0) {};
        \node[hasse,label=right:{$\mathcal C(\mathcal{N}_{\mathcal{T}}(\mathcal{A})^\vee)$}] (1) at (0,1.5) {};
        \node[hasse,label=right:{$\mathcal C(\mathcal{N}_{\mathcal{T}}(\mathcal{D})^\vee)$}] (2) at (0,3) {};
        %\node[hasse] (2) at (0,3) {};
        \node[hasse,label=right:{$\mathcal C(\mathcal{T}^\vee)$}] (3) at (0,4.5) {};
        \draw (0)--(1) (1)--(2) (2)--(3);
        \node at (-0.5,3.75) {$c_2$};
        \node at (-0.5,2.25) {$a_{n-3}$};
        \node at (-0.5,0.75) {$a_{n-1}$};
        \node at (0,3.5) {};
        %\node at (0,-1) {$n\geq4$};
        \draw [thick,decoration={brace,mirror,raise=1cm},decorate] (3) -- (2) node [pos=0.5,anchor=north,xshift=-2cm,yshift=0.2cm] {$\mathcal C(\mathcal{D}^\vee)$}; 
        \draw [thick,decoration={brace,mirror,raise=1cm},decorate] (2) -- (1) node [pos=0.5,anchor=north,xshift=-2.2cm,yshift=0.2cm] {$\mathcal{C}(\mathcal{N}_{\mathcal{A}}(\mathcal{D})^\vee)$}; 
        \draw [thick,decoration={brace,mirror,raise=2.5cm},decorate] (1) -- (3) node [pos=0.5,anchor=north,xshift=3.5cm,yshift=0.2cm] {$\mathcal C(\mathcal{A}^\vee)$}; 
        %\draw [thick,decoration={brace,mirror,raise=.5cm},decorate] (3) -- (2) node [pos=0.5,anchor=north,xshift=-1.5cm,yshift=0.2cm] {$\mathcal C(\mathcal{D}^\vee)$}; 
    \end{tikzpicture}}}\qquad.
\label{eq_hassemirror}
\end{equation}

A further check can be performed using the monopole formula \cite{Cremonesi:2013lqa}:
\begin{equation}
\mathrm{HS}_\mathcal{C}=\sum_{m\in \Gamma_{\widehat G}^{\ast}/W_{\widehat G}}
\,t^{2\Delta(m)}\,P_G(t^2;m),
\end{equation}
where $\Delta(m)$ is the conformal dimension and $P_G(t;m)$ is the dressing factor. In particular, for $\mathcal{T}^\vee$, the conformal dimension is
\begin{align}
2\Delta(q,\{m_i\})
=&-2\sum_{i=1}^{n-1}
\left|m_{i,1}-m_{i,2}\right|
+\sum_{\alpha=1}^{2}
\left|q-2m_{1,\alpha}\right|
+\sum_{\alpha=1}^{2}
\left|q-2m_{n-1,\alpha}\right| \notag\\
&+\sum_{i=1}^{n-2}
\sum_{\alpha,\beta=1}^{2}
\left|m_{i,\alpha}-m_{i+1,\beta}\right|,
\label{eq_confd}
\end{align}
where $q$ is the magnetic flux for $\mathrm{U}(1)$ and $\{m_{i,1},m_{i,2}\}$ are the magnetic fluxes for $i$-th $\mathrm{U}(2)$. The asymmetry terms $\vert q-2m_{1,\alpha}\vert$ and $\vert q-2m_{n-1,\alpha}\vert$ are contributions of the non-simply laced edges.
The dressing factor is
\begin{equation}
P(t^2;p,\{m_i\})
=
P_{U(1)}(t;p)
\prod_{i=1}^{n-1} P_{U(2)}(t;m_i),
\end{equation}
where
\begin{equation}
P_{U(1)}(t^2;p)=\frac{1}{1-t^2},
\end{equation}
and
\begin{equation}
P_{U(2)}(t^2;m_i)
=
\begin{cases}
\dfrac{1}{(1-t^2)(1-t^4)}, 
& m_{i,1}=m_{i,2},\\
\dfrac{1}{(1-t^2)^2}, 
& m_{i,1}>m_{i,2}.
\end{cases}
\end{equation}

The unframed quiver $\mathcal{T}^\vee$ given in \eqref{eq_Tm} has an overall free $\mathrm{U}(1)$ factor. This free $\mathrm{U}(1)$ manifests in \eqref{eq_confd} has a shift symmetry, which can be removed by setting $m_{i,\alpha}=0$ for any fixed $i$ and $\alpha$\footnote{There are many other equivalent ways to remove the shift symmetry.}. An overall $\frac{1}{1-t^2}$ factor from the dressing factor, which corresponds to the Casimir of the free $\mathrm{U}(1)$, also needs to be removed.
For $n=3,4$, the results are:
\begin{equation}
    \mathrm{HS}_{\mathcal{C}^\vee}(3)=
\frac{1+5t^2+26t^4+55t^6+78t^8+55t^{10}+26t^{12}+5t^{14}+t^{16}}
{(1-t^2)^4(1-t^4)^4},
\end{equation}
\begin{align}
    \mathrm{HS}_{\mathcal{C}^\vee}(4)=
    \mathrm{PE}[6t^2+6t^4]&
    (1+10t^2+78t^4+332t^6+911t^8+1606t^{10}+1956t^{12}
\notag\\&+1606t^{14}+911t^{16}+332t^{18}+78t^{20}+10t^{22}+t^{24})
,
\end{align}
in perfect agreement with \eqref{eq_HS3} and \eqref{eq_HS4}. The same agreement holds for arbitrary $n$.

With both matching of Hasse diagrams and Hilbert series, it can be confirmed that:
\begin{equation}
    \mathcal{C}(\mathcal{T}^\vee) \cong \mathcal{H}(\mathcal{T}).
\label{eq_idCHT}
\end{equation}
Next, we turn to the more challenging task of computing a Higgs branch for a non-simply laced quiver.

\subsection{Symplectic duality and Higgs branch of $\mathcal{T}^\vee$}
\label{sec_symplecticdual}

%To claim $\mathcal{T}$ and $\mathcal{T}^\vee$ to be a mirror pair, $\mathcal{H}(\mathcal{T}^\vee)$ and $\mathcal{C}(\mathcal{T})$ need to be computed and equal. The difficulty is that $\mathcal{H}(\mathcal{T}^\vee)$ cannot be computed directly. However if there is a method to construct $\mathcal{C}(\mathcal{T})$ from $\mathcal{H}(\mathcal{T})$ using only local geometry, then $\mathcal{H}(\mathcal{T}^\vee)$ can also be constructed from $\mathcal{C}(\mathcal{T}^\vee)$ using the same method. In this case, if $\mathcal{H}(\mathcal{T})\cong\mathcal{C}(\mathcal{T}^\vee)$ then by construction $\mathcal{C}(\mathcal{T})\cong\mathcal{H}(\mathcal{T}^\vee)$.

The reconstruction of $\mathcal{H}(\mathcal{T}^\vee)$ from $\mathcal{C}(\mathcal{T}^\vee)$ proceeds via inversion of the Hasse diagram, motivated by symplectic duality \cite{Braden:2022qua,Kamnitzer:2022sym,Webster:2023dim}. This approach relies on the assumption that the local geometry of the moduli space is fully captured by a finite set of transverse slices associated to quiver gauge theories with well-defined Higgs and Coulomb branches.

From the Hasse diagram \eqref{eq_hassemirror}, the local geometry of $\mathcal{C}(\mathcal{T}^\vee)$ is determined by the following set of building-block theories:
\begin{equation}
    \mathcal{G}^\vee:=\{\mathcal{D}^\vee,\ \mathcal{A}^\vee,\ \mathcal{N}_{\mathcal{A}}(\mathcal{D})^\vee,\ \mathcal{N}_{\mathcal{T}}(\mathcal{D})^\vee,\ \mathcal{N}_{\mathcal{T}}(\mathcal{A})^\vee\}.
\end{equation}%Similarly, the local geometry of $\mathcal{C}(\mathcal{T}^\vee)$ is completely determined by:
%\begin{equation}
%    \mathcal{G}^\vee:=\{\mathcal{D}^\vee,\ \mathcal{A}^\vee,\ \mathcal{N}_{\mathcal{A}}(\mathcal{D})^\vee,\ \mathcal{N}_{\mathcal{T}}(\mathcal{D})^\vee,\ \mathcal{N}_{\mathcal{T}}(\mathcal{A})^\vee\}.
%\end{equation}
%These two sets of data is identified under 3d mirror symmetry piecewisely, $\mathcal{G}\cong\mathcal{G}^\vee$.
%These two sets of data are piecewise identified under 3d mirror symmetry, \(\mathcal{G} \cong \mathcal{G}^\vee\).
%It is already a strong enough evidence to claim the two theories to be 3d mirror.
%If there is a way to build the local geometry of $\mathcal{C}(\mathcal{T})$ from $\mathcal{G}$, then so should $\mathcal{H}(\mathcal{T}^\vee)$ from $\mathcal{G}^\vee$.
For theories with sufficiently good properties, the Hasse diagrams of the Higgs and Coulomb branches are related by an inversion operation \cite{Grimminger:2020dmg}. In such cases, the local geometry of the Coulomb branch is entirely determined by that of the Higgs branch. More generally, the relations between the Higgs and Coulomb branches of a given theory fall under the concept of symplectic duality in the mathematics literature \cite{Braden:2022qua,Kamnitzer:2022sym,Webster:2023dim}.

Starting with $n=3$, $\mathcal{D}^\vee$ is empty. Consequently,
\[
\mathcal{N}_{\mathcal{A}}(\mathcal{D})^\vee = \mathcal{A}^\vee,
\qquad
\mathcal{N}_{\mathcal{T}}(\mathcal{D})^\vee = \mathcal{T}^\vee \, .
\]

For any $n$, the Higgs branch of $\mathcal{A}^\vee$ is given by
\begin{equation}
    A_{2n-3} := \mathbb{C}^2 / \mathbb{Z}_{2n-2}.
\end{equation}
Setting $n=3$, this reduces to the $A_3$ singularity.

The Higgs branch of $\mathcal{N}_{\mathcal{T}}(\mathcal{A})^\vee$ is given by the $A_{n-1}$ singularity for any $n$; setting $n=3$, this becomes $A_2$.
The inversion of the Hasse diagram of $\mathcal{C}(\mathcal{T}^\vee)$ reverses the partial order on the leaves and replaces each slice with the corresponding Coulomb branch. This procedure yields the Hasse diagram of $\mathcal{H}(\mathcal{T}^\vee)$:
\begin{equation}
    \vcenter{\hbox{\begin{tikzpicture}
        \node[hasse,label=right:{$\varnothing$}] (0) at (0,0) {};
        \node[hasse,label=right:{$\mathcal C(\mathcal{N}_{\mathcal{T}}(\mathcal{A}))$}] (1) at (0,1.5) {};
        \node[hasse,label=right:{$\mathcal C(\mathcal{T}^\vee)$}] (2) at (0,3) {};
        \draw (0)--(1) (1)--(2);
        %\node at (-1,3.75) {$h_{(1,1)^\mathsf{T}}\cong c_2$};
        %\node at (-0.5,2.25) {$a_{n-3}$};
        %\node at (-0.5,0.75) {$a_{n-1}$};
        \node at (0,3.5) {};
        \node at (-0.7,2.25) {$\bar h_{(1,1,2)^\mathsf{T}}$};
        \node at (-0.5,0.75) {$a_2$};
        %\node at (0,-1) {$n\geq4$};
        \draw [thick,decoration={brace,mirror,raise=1.5cm},decorate] (2) -- (1) node [pos=0.5,anchor=north,xshift=-2.5cm,yshift=0.2cm] {$\mathcal C(\mathcal{A}^\vee)$}; 
    \end{tikzpicture}}}
    \hspace{1cm} \xleftrightarrow{\text{Inversion}} \hspace{1cm}
    \vcenter{\hbox{\begin{tikzpicture}
        \node[hasse,label=right:{$\mathcal H(\mathcal{T}^\vee)$}] (0) at (0,0) {};
        \node[hasse,label=right:{$\mathcal H(\mathcal{N}_{\mathcal{T}}(\mathcal{A})^\vee)$}] (1) at (0,1.5) {};
        \node[hasse,label=right:{$\varnothing$}] (2) at (0,3) {};
        \draw (0)--(1) (1)--(2);
        \draw [thick,decoration={brace,mirror,raise=1cm},decorate] (1) -- (0) node [pos=0.5,anchor=north,xshift=-2cm,yshift=0.2cm] {$\mathcal H(\mathcal{A}^\vee)$}; 
        \node at (-0.5,2.25) {$A_2$};
        \node at (-0.5,0.75) {$A_3$};
    \end{tikzpicture}}}\qquad.
    \label{eq_invertn3}
\end{equation}
This inversion can also be done in the reverse direction from $\mathcal{H}(\mathcal{T}^\vee)$ to $\mathcal{C}(\mathcal{T}^\vee)$.

For $n\geq4$, there is a subtlety to apply inversion to $\mathcal{C}(\mathcal{T}^\vee)$, which can already be seen from the inversion of $\mathcal{C}(\mathcal{A}^\vee)$ and $\mathcal{C}(\mathcal{D}^\vee)$. The Hasse diagram of $\mathcal{C}(\mathcal{A}^\vee)$ has two slices, while the Hasse diagram of $\mathcal{H}(\mathcal{A}^\vee)$ has only one slice. Similarly, the Hasse diagram of $\mathcal{C}(\mathcal{D}^\vee)$ has one slice, while the Hasse diagram of $\mathcal{H}(\mathcal{D}^\vee)$ has no slices, as its Higgs branch geometry is trivial.
The na\"ive inversion still works for $\mathcal{D}^\vee$, the leaves $\varnothing$ and $\mathcal{C}(\mathcal{D})^\vee$ are identified after inversion into a single leaf, and the Coulomb branch geometry $c_2$ is replaced by a trivial Higgs branch:
\begin{equation}
\vcenter{\hbox{\begin{tikzpicture}
        \node[hasse,label=right:{$\varnothing$}] (0) at (0,0) {};
        \node[hasse,label=right:{$\mathcal C(\mathcal{D}^\vee)$}] (1) at (0,1.5) {};
        \draw (0)--(1);
        %\draw [thick,decoration={brace,mirror,raise=1cm},decorate] (1) -- (0) node [pos=0.5,anchor=north,xshift=-2cm,yshift=0.2cm] {$\mathcal C(\mathcal{N}_{\mathcal{T}}(\mathcal{A}))$}; 
        %\node at (-0.7,2.25) {$h_{(1,1,2)^\mathsf{T}}$};
        \node at (-0.5,0.75) {$c_2$};
    \end{tikzpicture}}}
    \hspace{1cm} \xleftrightarrow{\text{Inversion}} \hspace{1cm}
    \vcenter{\hbox{\begin{tikzpicture}
        \node[hasse,label=right:{$\varnothing\cong\mathcal{H}(\mathcal{D}^\vee)$}] (0) at (0,0) {};
        %\node[hasse,label=right:{$\mathcal C(\mathcal{A})$}] (1) at (0,1.5) {};
        %\node[hasse,label=right:{$\mathcal C(\mathcal{T})$}] (2) at (0,3) {};
        %\draw (0)--(1) (1)--(2);
        %\node at (-1,3.75) {$h_{(1,1)^\mathsf{T}}\cong c_2$};
        %\node at (-0.5,2.25) {$a_{n-3}$};
        %\node at (-0.5,0.75) {$a_{n-1}$};
        %\node at (0,3.5) {};
        %\node at (-0.5,2.25) {$A_2$};
        %\node at (-0.5,0.75) {$A_3$};
        %\node at (0,-1) {$n\geq4$};
        %\draw [thick,decoration={brace,mirror,raise=1cm},decorate] (2) -- (1) node [pos=0.5,anchor=north,xshift=-2cm,yshift=0.2cm] {$\mathcal C(\mathcal{N}_{\mathcal{T}}(\mathcal{A}))$}; 
    \end{tikzpicture}}}\qquad.
\end{equation}
If one keeps in mind that $\varnothing \cong \mathcal{H}(\mathcal{D}^\vee)$ on the Higgs branch, the inversion can also be performed in the reverse direction, from the Higgs branch to the Coulomb branch. 

However, if one applies the inversion to $\mathcal{C}(\mathcal{A}^\vee)$ slice by slice, one obtains:
\begin{equation}
\vcenter{\hbox{\begin{tikzpicture}
        \node[hasse,label=right:{$\varnothing$}] (0) at (0,0) {};
        \node[hasse,label=right:{$\mathcal C(\mathcal{N}_{\mathcal{A}}(\mathcal{D})^\vee)$}] (1) at (0,1.5) {};
        \node[hasse,label=right:{$\mathcal C(\mathcal{A}^\vee)$}] (2) at (0,3) {};
        \draw (0)--(1) (1)--(2);
        \draw [thick,decoration={brace,mirror,raise=1cm},decorate] (2) -- (1) node [pos=0.5,anchor=north,xshift=-2cm,yshift=0.2cm] {$\mathcal C(\mathcal{D}^\vee)$}; 
        \node at (-0.5,2.25) {$c_2$};
        \node at (-0.5,0.75) {$a_{n-3}$};
    \end{tikzpicture}}}
    \hspace{1cm} \xrightarrow{\text{Na\"ive inversion}} \hspace{1cm}
    \vcenter{\hbox{\begin{tikzpicture}
        \node[hasse,label=right:{$\mathcal H(\mathcal{A}^\vee)$}] (0) at (0,0) {};
        \node[hasse,label=right:{$\varnothing\cong\mathcal{H}(\mathcal{D}^\vee)$}] (1) at (0,1.5) {};
        %\node[hasse,label=right:{$\mathcal C(\mathcal{T})$}] (2) at (0,3) {};
        \draw (0)--(1);
        %\node at (-1,3.75) {$h_{(1,1)^\mathsf{T}}\cong c_2$};
        %\node at (-0.5,2.25) {$a_{n-3}$};
        %\node at (-0.5,0.75) {$a_{n-1}$};
        \node at (0,3.5) {};
        %\node at (-0.5,2.25) {$A_2$};
        \node at (-0.5,0.75) {$A_{n-3}$};
        %\node at (0,-1) {$n\geq4$};
        %\draw [thick,decoration={brace,mirror,raise=1cm},decorate] (2) -- (1) node [pos=0.5,anchor=north,xshift=-2cm,yshift=0.2cm] {$\mathcal C(\mathcal{N}_{\mathcal{T}}(\mathcal{A}))$};
    \end{tikzpicture}}}\qquad,
\end{equation}
which is obviously wrong, as the Higgs branch geometry should be $A_{2n-3}$ instead of $A_{n-3}$. The point is, for this case, even if the Higgs branch geometry of $\mathcal{D}^\vee$ is trivial, the Higgs branch of $\mathcal{A}^\vee$ is not simply given by $\mathcal{H}(\mathcal{N}_{\mathcal{A}}(\mathcal{D})^\vee)$.

The role of $\mathcal{D}^\vee$ illustrates an important subtlety: even when a theory has a trivial Higgs branch, it contributes nontrivially to the structure of the full moduli space through deformation parameters such as FI terms. Consequently, inversion is not applied at the level of individual slices, but rather on suitable building blocks that retain global information.

The correct inversion is:
\begin{equation}
\vcenter{\hbox{\begin{tikzpicture}
        \node[hasse,label=right:{$\varnothing$}] (0) at (0,0) {};
        \node[hasse,label=right:{$\mathcal C(\mathcal{N}_{\mathcal{A}}(\mathcal{D})^\vee)$}] (1) at (0,1.5) {};
        \node[hasse,label=right:{$\mathcal C(\mathcal{A}^\vee)$}] (2) at (0,3) {};
        \draw (0)--(1) (1)--(2);
        %\draw [thick,decoration={brace,mirror,raise=1cm},decorate] (2) -- (1) node [pos=0.5,anchor=north,xshift=-2cm,yshift=0.2cm] {$\mathcal C(\mathcal{D}^\vee)$}; 
        \node at (-0.5,2.25) {$c_2$};
        \node at (-0.5,0.75) {$a_{n-3}$};
    \end{tikzpicture}}}
    \hspace{1cm} \xleftrightarrow{\text{Inversion}} \hspace{1cm}
    \vcenter{\hbox{\begin{tikzpicture}
        \node[hasse,label=right:{$\mathcal H(\mathcal{A}^\vee)$}] (0) at (0,0) {};
        \node[hasse,label=right:{$\varnothing\cong\mathcal{H}(\mathcal{D}^\vee)$}] (1) at (0,1.5) {};
        %\node[hasse,label=right:{$\mathcal C(\mathcal{T})$}] (2) at (0,3) {};
        \draw (0)--(1);
        %\node at (-1,3.75) {$h_{(1,1)^\mathsf{T}}\cong c_2$};
        %\node at (-0.5,2.25) {$a_{n-3}$};
        %\node at (-0.5,0.75) {$a_{n-1}$};
        \node at (0,3.5) {};
        %\node at (-0.5,2.25) {$A_2$};
        \node at (-0.6,0.75) {$A_{2n-3}$};
        %\node at (0,-1) {$n\geq4$};
        %\draw [thick,decoration={brace,mirror,raise=1cm},decorate] (2) -- (1) node [pos=0.5,anchor=north,xshift=-2cm,yshift=0.2cm] {$\mathcal C(\mathcal{N}_{\mathcal{T}}(\mathcal{A}))$};
    \end{tikzpicture}}}\qquad.
\end{equation}
It may seem counterintuitive that a trivial geometry contributes nontrivially to the Higgs branch; however, the data of $\mathcal{D}^\vee$ is indeed encoded in the Higgs branch through FI deformations. The details are discussed in Section~\ref{sec_HTinvert}.
To apply inversion on $\mathcal{C}(\mathcal{T}^\vee)$, the fundamental building block should be $\mathcal{A}^\vee$ and $\mathcal{N}_{\mathcal{T}}(\mathcal{A})^\vee$ instead of $\mathcal{D}^\vee$, $\mathcal{N}_{\mathcal{A}}(\mathcal{D})^\vee$, and $\mathcal{N}_{\mathcal{T}}(\mathcal{D})^\vee$. The inversion is as follows:
\begin{equation}
    \vcenter{\hbox{\begin{tikzpicture}
        \node[hasse,label=right:{$\varnothing$}] (0) at (0,0) {};
        \node[hasse,label=right:{$\mathcal C(\mathcal{N}_{\mathcal{T}}(\mathcal{A})^\vee)$}] (1) at (0,1.5) {};
        \node[hasse,label=right:{$\mathcal C(\mathcal{N}_{\mathcal{T}}(\mathcal{D})^\vee)$}] (2) at (0,3) {};
        \node[hasse,label=right:{$\mathcal C(\mathcal{T}^\vee)$}] (3) at (0,4.5) {};
        \draw (0)--(1) (1)--(2) (2)--(3);
        \node at (-0.5,3.75) {$c_2$};
        \node at (-0.5,2.25) {$a_{n-3}$};
        \node at (-0.5,0.75) {$a_{n-1}$};
        \draw [thick,decoration={brace,mirror,raise=2.5cm},decorate] (1) -- (3)
        node [pos=0.5,anchor=north,xshift=3.2cm,yshift=0.2cm]
        {$\mathcal{C}(\mathcal{A}^\vee)$};
        %\draw [thick,decoration={brace,mirror,raise=1cm},decorate] (2) -- (0)node [pos=0.5,anchor=north,xshift=-2cm,yshift=0.2cm]{$\overline{\mathcal{O}^\mathrm{n.min}_{\mathfrak{sl}_{n}}}$};
    \end{tikzpicture}}}
    \hspace{0.3cm} \xleftrightarrow{\text{Inversion}} \hspace{0.3cm}
    \vcenter{\hbox{\begin{tikzpicture}
        \node[hasse,label=right:{$\mathcal{H}(\mathcal{T}^\vee)$}] (0) at (0,0) {};
        \node[hasse,label=right:{$\mathcal{H}(\mathcal{N}_{\mathcal{T}}(\mathcal{A})^\vee)\times\mathcal{H}(\mathcal{D^\vee})$}] (1) at (0,1.5) {};
        \node[hasse,label=right:{$\varnothing$}] (2) at (0,3) {};
        \draw (0)--(1) (1)--(2);
        %\node at (-1,3.75) {$h_{(1,1)^\mathsf{T}}\cong c_2$};
        %\node at (-0.5,2.25) {$a_{n-3}$};
        %\node at (-0.5,0.75) {$a_{n-1}$};
        \node at (0,3.5) {};
        \node at (-0.5,2.25) {$A_{n-1}$};
        \node at (-0.6,0.75) {$A_{2n-3}$};
        %\node at (0,-1) {$n\geq4$};
        \draw [thick,decoration={brace,mirror,raise=1.3cm},decorate] (1) -- (0) node [pos=0.5,anchor=north,xshift=-2.2cm,yshift=0.2cm] {$\mathcal H(\mathcal{A}^\vee)$}; 
    \end{tikzpicture}}}\qquad.
    \label{eq_invertn4}
\end{equation}
This inversion provides the first example in which the Higgs branch of a non-simply laced theory can be computed without explicit knowledge of its 3d mirror theory. A similar method can be applied to any non-simply laced quiver whose Coulomb branch local geometry is determined by theories with well-defined Higgs and Coulomb branches.

\subsection{Coulomb branch of $\mathcal{T}$ and Mirror symmetry}
\label{sec_HTinvert}
The Coulomb branch of $\mathcal{T}$ is well defined. Its Hilbert series can be computed via monopole formula. For $n=3,4$, the results are
\begin{equation}
    \mathrm{HS_\mathcal{C}(3)}=\frac{1+2t^4+4t^6+2t^8+t^{12}}
{(1-t^2)(1-t^4)^2(1-t^6)}.%=\mathrm{HS_{\mathcal{T}^\vee}(3)}.
\end{equation}
\begin{equation}
    \mathrm{HS_\mathcal{C}(4)}=\frac{
1+t^2+t^4+2t^6+6t^8+6t^{10}+6t^{12}
+2t^{14}+t^{16}+t^{18}+t^{20}
}
{
(1-t^4)(1-t^6)^2(1-t^8)
}.%=\mathrm{HS_{\mathcal{T}^\vee}(4)}.
\end{equation}
The Hasse diagram of $\mathcal{C}(\mathcal{T})$ can, in principle, be computed explicitly. However, due to the presence of both $\mathrm{SU}(2)$ and $\mathrm{U}(1)$ gauge nodes, the computation is difficult in practice. Moreover, quiver contraction techniques for such Coulomb branches are not yet well developed. Consequently, for $\mathcal{C}(\mathcal{T})$, the most practical way to obtain its Hasse diagram is via inversion, using the known stratification of the Higgs branch.

The inversion of the Hasse diagram of $\mathcal{H}(\mathcal{T})$ reverses the partial order on the leaves and replaces each slice with the corresponding Coulomb branch.
For $n=3$,
\begin{equation}
\vcenter{\hbox{\begin{tikzpicture}
        \node[hasse,label=right:{$\mathcal H(\mathcal{T})$}] (0) at (0,0) {};
        \node[hasse,label=right:{$\mathcal H(\mathcal{A})$}] (1) at (0,1.5) {};
        \node[hasse,label=right:{$\varnothing$}] (2) at (0,3) {};
        \draw (0)--(1) (1)--(2);
        \draw [thick,decoration={brace,mirror,raise=1cm},decorate] (1) -- (0) node [pos=0.5,anchor=north,xshift=-2cm,yshift=0.2cm] {$\mathcal C(\mathcal{N}_{\mathcal{T}}(\mathcal{A}))$}; 
        \node at (-0.7,2.25) {$\bar h_{(1,1,2)^\mathsf{T}}$};
        \node at (-0.5,0.75) {$a_2$};
    \end{tikzpicture}}}
    \hspace{1cm} \xleftrightarrow{\text{Inversion}} \hspace{1cm}
    \vcenter{\hbox{\begin{tikzpicture}
        \node[hasse,label=right:{$\varnothing$}] (0) at (0,0) {};
        \node[hasse,label=right:{$\mathcal C(\mathcal{A})$}] (1) at (0,1.5) {};
        \node[hasse,label=right:{$\mathcal C(\mathcal{T})$}] (2) at (0,3) {};
        \draw (0)--(1) (1)--(2);
        %\node at (-1,3.75) {$h_{(1,1)^\mathsf{T}}\cong c_2$};
        %\node at (-0.5,2.25) {$a_{n-3}$};
        %\node at (-0.5,0.75) {$a_{n-1}$};
        \node at (0,3.5) {};
        \node at (-0.5,2.25) {$A_2$};
        \node at (-0.5,0.75) {$A_3$};
        %\node at (0,-1) {$n\geq4$};
        \draw [thick,decoration={brace,mirror,raise=1cm},decorate] (2) -- (1) node [pos=0.5,anchor=north,xshift=-2cm,yshift=0.2cm] {$\mathcal C(\mathcal{N}_{\mathcal{T}}(\mathcal{A}))$}; 
    \end{tikzpicture}}}\qquad.
\end{equation}

For $n\geq4$,
\begin{equation}
    \vcenter{\hbox{\begin{tikzpicture}
        \node[hasse,label=right:{$\mathcal H(\mathcal{T})$}] (0) at (0,0) {};
        \node[hasse,label=right:{$\mathcal H(\mathcal{A})$}] (1) at (0,1.5) {};
        \node[hasse,label=right:{$\mathcal H(\mathcal{D})$}] (2) at (0,3) {};
        \node[hasse,label=right:{$\varnothing$}] (3) at (0,4.5) {};
        \draw (0)--(1) (1)--(2) (2)--(3);
        \node at (-0.5,3.75) {$c_2$};
        \node at (-0.5,2.25) {$a_{n-3}$};
        \node at (-0.5,0.75) {$a_{n-1}$};
        %\draw [thick,decoration={brace,mirror,raise=1.5cm},decorate] (1) -- (3) node [pos=0.5,anchor=north,xshift=2.2cm,yshift=0.2cm] {$\mathcal{H}(\mathcal{A})$};
        \draw [thick,decoration={brace,mirror,raise=1.5cm},decorate] (0) -- (1)
        node [pos=0.5,anchor=north,xshift=2.6cm,yshift=0.2cm]
        {$\mathcal{H}(\mathcal{N}_{\mathcal{T}}(\mathcal{A}))$};
        %\draw [thick,decoration={brace,mirror,raise=1cm},decorate] (2) -- (0)node [pos=0.5,anchor=north,xshift=-2cm,yshift=0.2cm]{$\overline{\mathcal{O}^\mathrm{n.min}_{\mathfrak{sl}_{n}}}$};
    \end{tikzpicture}}}
    \hspace{0.5cm} \xleftrightarrow{\text{Inversion}} \hspace{0.5cm}
    \vcenter{\hbox{\begin{tikzpicture}
        \node[hasse,label=right:{$\varnothing\cong\mathcal{C}(\mathcal{D})$}] (0) at (0,0) {};
        \node[hasse,label=right:{$\mathcal C(\mathcal{A})$}] (1) at (0,1.5) {};
        \node[hasse,label=right:{$\mathcal C(\mathcal{T})$}] (2) at (0,3) {};
        \draw (0)--(1) (1)--(2);
        %\node at (-1,3.75) {$h_{(1,1)^\mathsf{T}}\cong c_2$};
        %\node at (-0.5,2.25) {$a_{n-3}$};
        %\node at (-0.5,0.75) {$a_{n-1}$};
        \node at (0,3.5) {};
        \node at (-0.5,2.25) {$A_{n-1}$};
        \node at (-0.6,0.75) {$A_{2n-3}$};
        %\node at (0,-1) {$n\geq4$};
        \draw [thick,decoration={brace,mirror,raise=1cm},decorate] (2) -- (1) node [pos=0.5,anchor=north,xshift=-2cm,yshift=0.2cm] {$\mathcal C(\mathcal{N}_{\mathcal{T}}(\mathcal{A}))$}; 
    \end{tikzpicture}}}\qquad.
\end{equation}
Note that setting $n=3$ in the general $n$ diagram agrees with the special case $n=3$.
The resulting Hasse diagrams match with $\mathcal{H}(\mathcal{T}^\vee)$ in \eqref{eq_invertn3} and \eqref{eq_invertn4}, which verifies the conjecture that
\begin{equation}
    \mathcal{C}(\mathcal{T}) \cong \mathcal{H}(\mathcal{T}^\vee).
\label{eq_cthtm}
\end{equation}
Note that even if $\mathcal{H}(\mathcal{T}) \cong \mathcal{C}(\mathcal{T}^\vee)$, their inversion can be different, \eqref{eq_cthtm} is not automatically true.
The inversion depends on the local description of leaves and slices, so it is necessary to apply inversion on both $\mathcal{H}(\mathcal{T})$ and $\mathcal{C}(\mathcal{T}^\vee)$ .

With \eqref{eq_idCHT} and \eqref{eq_cthtm}, there are sufficiently strong evidence to claim $\mathcal{T}$ and $\mathcal{T}^\vee$ are related by 3d mirror symmetry,
\begin{equation}
    \mathcal{T} \xleftrightarrow{\text{mirror}} \mathcal{T}^\vee.
\end{equation}

\subsection{Mass deformations and FI deformations}

Recall that in $\mathcal{T}^\vee$, $\mathcal{H}(\mathcal{D}^\vee)$ is trivial, yet it contributes nontrivially to the Hasse diagrams of $\mathcal{H}(\mathcal{A}^\vee)$ and $\mathcal{H}(\mathcal{T}^\vee)$. A similar phenomenon occurs for $\mathcal{T}$: although $\mathcal{C}(\mathcal{D})$ is trivial, it contributes nontrivially to $\mathcal{C}(\mathcal{A})$ and $\mathcal{C}(\mathcal{T})$.

% The contribution of $\mathcal{D}$ can be seen explicitly through mass deformation of $\mathcal{A}$ \eqref{eq_A}: give masses to charge $1$ hypers and integrate them out, the remaining low energy theory is a $\mathrm{U}(1)$ theory with $n\!-2\!$ charge $2$ hypers. The magnetic flux can only breaks the symmetry to a $\mathbb{Z}_2$ residual subgroup. This $\mathbb{Z}_2$ is precisely the difference between charge $1$ and charge $2$ weight lattice. If the lattice get rescaled\footnote{This operation corresponds to a gauging of a $\mathbb{Z}_2$ electric $1$-form symmetry, see \cite{Gaiotto:2014kfa,Grimminger:2025fgj}.} to weight $1$, one obtain the $\mathcal{N}_{\mathcal{A}}(\mathcal{D})$ \eqref{eq_NAD} theory and then the resulting Coulomb branch is $A_{n-3}$ singularity, which is expected from the na\"ive inversion. The charge $2$ hypers are charged trivially under this $\mathbb{Z}_2$. The charge $1$ hypers are charged non-trivially under this $\mathbb{Z}_2$, and gives theory $\mathcal{D}$ \eqref{eq_D}.
The contribution of $\mathcal{D}$ can be seen explicitly through a mass deformation of $\mathcal{A}$~\eqref{eq_A}: giving masses to the charge-$1$ hypermultiplets and integrating them out, the remaining low-energy theory is a $\mathrm{U}(1)$ gauge theory with $n-2$ charge-$2$ hypermultiplets. The magnetic fluxes can only break the gauge symmetry to a residual $\mathbb{Z}_2$ subgroup. This $\mathbb{Z}_2$ precisely captures the distinction between the charge-$1$ and charge-$2$ weight lattices. 

If the lattice is rescaled\footnote{This operation corresponds to gauging a $\mathbb{Z}_2$ electric one-form symmetry; see \cite{Gaiotto:2014kfa,Grimminger:2025fgj}.} to unit charge, one obtains the theory $\mathcal{N}_{\mathcal{A}}(\mathcal{D})$~\eqref{eq_NAD}, whose Coulomb branch is the $A_{n-3}$ singularity, in agreement with the naive inversion. 

The charge-$2$ hypermultiplets are neutral under this $\mathbb{Z}_2$, while the charge-$1$ hypermultiplets transform nontrivially under it and give rise to the theory $\mathcal{D}$~\eqref{eq_D}.

A 3d mirror dual procedure can be applied to $\mathcal{A}^\vee$ by turning on certain FI terms.

In summary, while the construction of $\mathcal{T}^\vee$ is not derived from a first-principles Lagrangian, the combined evidence from Hilbert series, Hasse diagrams, and consistency with symplectic duality provides strong support for the proposed mirror pair.

\section{Discrete quotient from $D$ to $A$}
\label{sec_4}

% This section presents an interesting quotient relation on the Coulomb branches of non-simply laced quivers. In Section \ref{sec_nslm}, the decoupling of a overall free $\mathrm{U}(1)$ from $\mathcal{T}^\vee$ is discussed, which can be done by setting $m_{i,\alpha}=0$. If instead, set $q=0$, the $\mathrm{U}(1)$ gauge node is turned into a flavour node, and the resulting theory has a non-simply laced quiver:
In the previous sections, the non-simply laced quiver \(\mathcal{T}^\vee\) was shown to reproduce $\overline{T^*\mathcal{O}^{\mathrm{min}}_{\mathfrak{sl}_n}}^{\mathrm{aff}}$. In this section we study a closely related quotient construction, related to different choices of ungauging schemes \cite{Hanany:2020jzl}.
In Section~\ref{sec_nslm}, the decoupling of an overall free $\mathrm{U}(1)$ factor from $\mathcal{T}^\vee$ is discussed, which can be achieved by setting one magnetic charge $m_{i,\alpha}=0$. 

If instead one sets $q=0$, the $\mathrm{U}(1)$ gauge node is effectively converted into a flavour node, and the resulting theory is described by a non-simply laced quiver:
\begin{equation}
    \mathcal{T}^\vee_1=
    \raisebox{-0.5\height}{\begin{tikzpicture}
    \node[gauge,label=below:{2}] (0) at (0,0) {};
    \node[gauge,label=below:{2}] (1) at (1,0) {};
    \node (2) at (2,0) {$\cdots$};
    \node[gauge,label=below:{2}] (3) at (3,0) {};
    \node[gauge,label=below:{2}] (4) at (4,0) {};
    %\node[gauge,label=below:{1}] (5) at (5,0) {};
    \node[flavour,label=above:{1}] (4a) at (4,1) {};
    %\node[flavour,label=above:{1}] (3a) at (3,1) {};
    %\node[gauge,label=above:{1}] (2a) at (2,1) {};
    %\node[flavour,label=above:{1}] (1a) at (1,1) {};
    \node[flavour,label=above:{1}] (0a) at (0,1) {};
    \draw (0)--(1)--(2)--(3)--(4);
    \draw[transform canvas={xshift=-1.5pt}] (0)--(0a) (4)--(4a);
    \draw[transform canvas={xshift=1.5pt}] (0)--(0a) (4)--(4a);
    \draw (-0.2,0.4)--(0,0.6)--(0.2,0.4) (3.8,0.4)--(4,0.6)--(4.2,0.4);
    %\draw (0.8,1.1)--(1.1,1.1)--(1.1,0.8) (3.2,1.1)--(2.9,1.1)--(2.9,0.8);
    %\draw[shift={(1.1,0.55)},rotate=26.565] (-0.212,0.212)--(0,0)--(-0.212,-0.212);
    %\draw[shift={(2.9,0.55)},rotate=153.435](-0.212,0.212)--(0,0)--(-0.212,-0.212);
    %\draw (0a) .. controls (4,1) .. (4);
    \draw [thick,decoration={brace,mirror,raise=0.8cm},decorate] (0) -- (4) 
node [pos=0.5,anchor=north,yshift=-0.85cm] {$n-1$}; 
    %\draw (-.2,.6)--(0,0.4)--(.2,.6);
     %\node at (3.5,.5) {$k$};
     \end{tikzpicture}}.
\end{equation}
However, setting $q=0$ not only removes the free $\mathrm{U}(1)$ factor, but also rescales the magnetic lattice by a $\mathbb{Z}_2$ factor. In other words, the magnetic lattice of $\mathcal{T}^\vee$, denoted by $\Gamma$, and that of $\mathcal{T}^\vee_1$, denoted by $\Gamma_1$, are related as follows:

\begin{equation}
    \Gamma/\Gamma_1 \cong \mathbb{Z}\times\mathbb{Z}_2.
\end{equation}
As a consequence, the Coulomb branch of $\mathcal{T}_1^\vee$ is a $\mathbb{Z}_2$ quotient of the Coulomb branch of $\mathcal{T}^\vee$,

\begin{equation}
    \mathcal{C}(\mathcal{T}^\vee_1)\cong\mathcal{C}(\mathcal{T}^\vee)/\mathbb{Z}_2.
\label{eq_idT1T}
\end{equation}
This $\mathbb{Z}_2$ quotient on the Coulomb branch arises from gauging a $\mathbb{Z}_2$ subgroup of the magnetic $\mathrm{U}(1)$ symmetry associated with the $\mathrm{U}(1)$ node in $\mathcal{T}^\vee$. The Higgs branch, however, remains unchanged under such discrete gauging of the magnetic symmetry,

\begin{equation}
    \mathcal{H}(\mathcal{T}_1^\vee)\cong \mathcal{H}(\mathcal{T}^\vee).
\end{equation}

To study $\mathcal{C}(\mathcal{T}^\vee_1)$, it is helpful to define a new quiver $\mathcal{T}^\vee_2$ as follows
\begin{equation}
    \mathcal{T}^\vee_2=
    \raisebox{-0.5\height}{\begin{tikzpicture}
    \node[gauge,label=below:{2}] (0) at (0,0) {};
    \node[gauge,label=below:{2}] (1) at (1,0) {};
    \node (2) at (2,0) {$\cdots$};
    \node[gauge,label=below:{2}] (3) at (3,0) {};
    \node[gauge,label=below:{2}] (4) at (4,0) {};
    %\node[gauge,label=below:{1}] (5) at (5,0) {};
    \node[flavour,label=above:{2}] (4a) at (4,1) {};
    %\node[flavour,label=above:{1}] (3a) at (3,1) {};
    %\node[gauge,label=above:{1}] (2a) at (2,1) {};
    %\node[flavour,label=above:{1}] (1a) at (1,1) {};
    \node[flavour,label=above:{2}] (0a) at (0,1) {};
    \draw (0)--(1)--(2)--(3)--(4);
    \draw (0)--(0a) (4)--(4a);
    %\draw (-0.2,0.4)--(0,0.6)--(0.2,0.4) (3.8,0.4)--(4,0.6)--(4.2,0.4);
    %\draw (0.8,1.1)--(1.1,1.1)--(1.1,0.8) (3.2,1.1)--(2.9,1.1)--(2.9,0.8);
    %\draw[shift={(1.1,0.55)},rotate=26.565] (-0.212,0.212)--(0,0)--(-0.212,-0.212);
    %\draw[shift={(2.9,0.55)},rotate=153.435](-0.212,0.212)--(0,0)--(-0.212,-0.212);
    %\draw (0a) .. controls (4,1) .. (4);
    \draw [thick,decoration={brace,mirror,raise=0.8cm},decorate] (0) -- (4) 
node [pos=0.5,anchor=north,yshift=-0.85cm] {$n-1$}; 
    %\draw (-.2,.6)--(0,0.4)--(.2,.6);
     %\node at (3.5,.5) {$k$};
     \end{tikzpicture}}.
\end{equation}
It is easy to see $\mathcal{T}^\vee_1$ and $\mathcal{T}^\vee_2$ have the same Coulomb branch,
\begin{equation}
    \mathcal{C}(\mathcal{T}^\vee_1)\cong\mathcal{C}(\mathcal{T}^\vee_2).
\label{eq_idT1T2}
\end{equation}
$\mathcal{T}^\vee_2$ is a balanced quiver of type $A_{n-1}$, as studied in \cite{Braverman:2016pwk,Bourget:2021siw}, its Coulomb branch is a slice in affine Grassmannian of $A_{n-1}$:
\begin{equation}
    \mathcal{C}(\mathcal{T}^\vee_2)\cong \overline{[\mathcal W_{A_{n-1}}]}_{0}^{2\omega_1+2\omega_{n-1}},
\end{equation}
where $\omega_i$ is the $i$-th fundamental coweight. Together with \eqref{eq_idHdc}, \eqref{eq_HTaffine}, \eqref{eq_idCHT}, \eqref{eq_idT1T}, and \eqref{eq_idT1T2},
\begin{equation}
    \overline{[\mathcal W_{A_{n-1}}]}_{0}^{2\omega_1+2\omega_{n-1}}\cong \overline{T^*\mathcal{O}^\mathrm{min}_{\mathfrak{sl}_n}}^{\mathrm{aff}}/\mathbb{Z}_2 \cong (d_{n+1}///\mathbb C^\times)/\mathbb{Z}_2.
\end{equation}
This result shows that the singularity $\overline{T^*\mathcal{O}^\mathrm{min}_{\mathfrak{sl}_n}}^{\mathrm{aff}}$ constructed in \cite{Fu:2026con} is a double cover of the slice in the affine Grassmannian of $A_{n-1}$ associated with twice the coweight of the adjoint representation.

$\mathcal{C}(\mathcal{T}^\vee_2)$ and $\mathcal{C}(\mathcal{T}^\vee_1)$ have the same unlabeled Hasse diagram:
\begin{equation}
        \vcenter{\hbox{\begin{tikzpicture}
        \node[hasse] (0) at (0,0) {};
        \node[hasse] (1) at (0,1.5) {};
        \node[hasse] (2l) at (-1,3) {};
        \node[hasse] (2r) at (1,3) {};
        \node[hasse] (3) at (0,4.5) {};
        \draw (0)--(1) (2r)--(1)--(2l) (2r)--(3)--(2l);
        \node at (-0.8,3.75) {$A_1$};
        \node at (0.8,3.75) {$A_1$};
        \node at (-0.8,2.25) {$A_2$};
        \node at (0.8,2.25) {$A_2$};
        \node at (-0.5,0.75) {$a_2$};
        \node at (0,-1) {$n=3$};
    \end{tikzpicture}}}
    \hspace{3cm}
    \vcenter{\hbox{\begin{tikzpicture}
        \node[hasse] (0) at (0,0) {};
        \node[hasse] (1) at (0,1.5) {};
        \node[hasse] (2) at (0,3) {};
        \node[hasse] (3l) at (-1,4.5) {};
        \node[hasse] (3r) at (1,4.5) {};
        \node[hasse] (4) at (0,6) {};
        \draw (0)--(1)--(2) (3r)--(2)--(3l) (3r)--(4)--(3l);
        \node at (-0.8,5.25) {$A_1$};
        \node at (0.8,5.25) {$A_1$};
        \node at (-0.8,3.75) {$A_1$};
        \node at (0.8,3.75) {$A_1$};
        \node at (-0.5,2.25) {$a_{n-3}$};
        \node at (-0.5,0.75) {$a_{n-1}$};
        \node at (0,-1) {$n=4$};
    \end{tikzpicture}}}\qquad,
\end{equation}
Comparing with \eqref{eq_hasseT}, one sees that the $\mathbb{Z}_2$ quotient acts only on the top slice, as

\begin{equation}
        \vcenter{\hbox{\begin{tikzpicture}
        \node[hasse] (0) at (0,0) {};
        \node[hasse] (1) at (0,1.5) {};
        \draw (0)--(1);
        \node at (-0.6,0.75) {$\bar h_{(1,1,2)}$};
    \end{tikzpicture}}}
    \hspace{1cm} \xrightarrow{\mathbb{Z}_2} \hspace{1cm}
    \vcenter{\hbox{\begin{tikzpicture}
        \node[hasse] (2) at (0,0) {};
        \node[hasse] (3l) at (-1,1.5) {};
        \node[hasse] (3r) at (1,1.5) {};
        \node[hasse] (4) at (0,3) {};
        \draw (3r)--(2)--(3l) (3r)--(4)--(3l);
        \node at (-0.8,2.25) {$A_1$};
        \node at (0.8,2.25) {$A_1$};
        \node at (-0.8,0.75) {$A_2$};
        \node at (0.8,0.75) {$A_2$};
    \end{tikzpicture}}}\qquad,
\end{equation}
\begin{equation}
        \vcenter{\hbox{\begin{tikzpicture}
        \node[hasse] (0) at (0,0) {};
        \node[hasse] (1) at (0,1.5) {};
        \draw (0)--(1);
        \node at (-0.5,0.75) {$c_2$};
    \end{tikzpicture}}}
    \hspace{1cm} \xrightarrow{\mathbb{Z}_2} \hspace{1cm}
    \vcenter{\hbox{\begin{tikzpicture}
        \node[hasse] (2) at (0,0) {};
        \node[hasse] (3l) at (-1,1.5) {};
        \node[hasse] (3r) at (1,1.5) {};
        \node[hasse] (4) at (0,3) {};
        \draw (3r)--(2)--(3l) (3r)--(4)--(3l);
        \node at (-0.8,2.25) {$A_1$};
        \node at (0.8,2.25) {$A_1$};
        \node at (-0.8,0.75) {$A_1$};
        \node at (0.8,0.75) {$A_1$};
    \end{tikzpicture}}}\qquad.
\end{equation}

Although the two theories have the same Coulomb branch, this does not imply that their Higgs branches are also the same. In fact:
\begin{equation}
    \mathcal{H}(\mathcal{T}^\vee_1)\neq\mathcal{H}(\mathcal{T}^\vee_2).
\end{equation}
As mentioned in Section~\ref{sec_HTinvert}, identical Coulomb branches do not necessarily lead to identical Higgs branches after inversion. The gauge-theoretic descriptions of the leaves and slices provide essential input for the inversion procedure, and these are often different even when two Coulomb branches correspond to the same symplectic singularity. 

In particular, in this case the gauge-theoretic descriptions of $\mathcal{C}(\mathcal{T}_1^\vee)$ and $\mathcal{C}(\mathcal{T}_2^\vee)$ differ, leading to distinct Higgs branch Hasse diagrams after inversion. The inversion should therefore be applied to a Hasse diagram in which each leaf and slice is properly labeled. Only after performing the inversion can one choose to suppress these labels for simplicity.

After applying inversion to $\mathcal{C}(\mathcal T^\vee_1)$, the unlabeled Hasse diagram of $\mathcal{H}(\mathcal T^\vee_1)$ is:
\begin{equation}
    \vcenter{\hbox{\begin{tikzpicture}
        \node[hasse] (0) at (0,0) {};
        \node[hasse] (1) at (0,1.5) {};
        \node[hasse] (2) at (0,3) {};
        \draw (0)--(1) (1)--(2);
        %\node at (-1,3.75) {$h_{(1,1)^\mathsf{T}}\cong c_2$};
        %\node at (-0.5,2.25) {$a_{n-3}$};
        %\node at (-0.5,0.75) {$a_{n-1}$};
        \node at (-0.5,2.25) {$A_{n-1}$};
        \node at (-0.6,0.75) {$A_{2n-3}$};
        %\node at (0,-1) {$n\geq4$};
        %\draw [thick,decoration={brace,mirror,raise=1cm},decorate] (2) -- (1) node [pos=0.5,anchor=north,xshift=-2cm,yshift=0.2cm] {$\mathcal C(\mathcal{N}_{\mathcal{T}}(\mathcal{A}))$};
        %\node at (0,-1) {$n=4$};
    \end{tikzpicture}}}\qquad.
\label{eq_hasset1h}
\end{equation}
It is the same as the Hasse diagram of $\mathcal{H}(\mathcal{T}^\vee)$, which agrees with the fact that gauging a discrete magnetic symmetry does not change the Higgs branch.
After applying inversion to $\mathcal{C}(\mathcal T^\vee_2)$, the unlabeled Hasse diagram of $\mathcal{H}(\mathcal T^\vee_2)$ is:
\begin{equation}
        \vcenter{\hbox{\begin{tikzpicture}
        %\node[hasse] (0) at (0,0) {};
        \node[hasse] (1) at (0,1.5) {};
        \node[hasse] (2l) at (-1,3) {};
        \node[hasse] (2r) at (1,3) {};
        \node[hasse] (3) at (0,4.5) {};
        \node[hasse] (4) at (0,6) {};
        \draw (3)--(4) (2r)--(1)--(2l) (2r)--(3)--(2l);
        \node at (-0.8,3.75) {$a_2$};
        \node at (0.8,3.75) {$a_2$};
        \node at (-0.8,2.25) {$a_1$};
        \node at (0.8,2.25) {$a_1$};
        \node at (-0.5,5.25) {$A_2$};
        \node at (0,0.5) {$n=3$};
    \end{tikzpicture}}}
    \hspace{3cm}
    \vcenter{\hbox{\begin{tikzpicture}
        %\node[hasse] (0) at (0,0) {};
        %\node[hasse] (1) at (0,1.5) {};
        \node[hasse] (2) at (0,3) {};
        \node[hasse] (3l) at (-1,4.5) {};
        \node[hasse] (3r) at (1,4.5) {};
        \node[hasse] (4) at (0,6) {};
        \node[hasse] (5) at (0,7.5) {};
        \node[hasse] (6) at (0,9) {};
        \draw (4)--(5)--(6) (3r)--(2)--(3l) (3r)--(4)--(3l);
        \node at (-0.8,5.25) {$a_1$};
        \node at (0.8,5.25) {$a_1$};
        \node at (-0.8,3.75) {$a_1$};
        \node at (0.8,3.75) {$a_1$};
        \node at (-0.5,6.75) {$A_{n-3}$};
        \node at (-0.5,8.25) {$A_{n-1}$};
        \node at (0,2) {$n\geq4$};
    \end{tikzpicture}}}\qquad.
    \label{eq_hasset2h}
\end{equation}
With \eqref{eq_hasset1h} and \eqref{eq_hasset2h}, it is clear that $\mathcal{H}(\mathcal T^\vee_1)$ and $\mathcal{H}(\mathcal T^\vee_2)$ are different.

The mirror theories of $\mathcal{T}^\vee_1$ and $\mathcal{T}^\vee_2$ are both well-defined theories.
The mirror theory of $\mathcal{T}^\vee_1$ is proposed to be:
\begin{equation}
    \mathcal{T}_1=\raisebox{-0.5\height}{\begin{tikzpicture}
    \node[gauge,label=below:{$\mathrm{SU}(2)$}] (0) at (0,0) {};
    \node[gauge,label=below:{$\mathrm{SO}(2)$}] (1) at (1.5,0) {};
    \node[flavour,label=above:{$\mathrm{O}(1)$}] (0al) at (-0.5,1) {};
    \node[gauge,label=above:{$\mathrm{O}(1)$}] (0ar) at (0.5,1) {};
    \draw (0al)--(0)--(0ar);
    \node at (0.75,0.3) {$n$};
    %\draw (0a) .. controls (4,1) .. (4);
    \draw[transform canvas={yshift=2pt}] (0)--(1);
    \draw[transform canvas={yshift=0pt}] (0)--(1);
    \draw[transform canvas={yshift=-2pt}] (0)--(1);
    %\draw [thick,decoration={brace,mirror,raise=0.8cm},decorate] (1) -- (4) node [pos=0.5,anchor=north,yshift=-0.85cm] {$n-2$}; 
    %\draw (-.2,.6)--(0,0.4)--(.2,.6);
     %\node at (3.5,.5) {$k$};
     \end{tikzpicture}},
\label{eq_T1}
\end{equation}
where $\mathrm{O}(1)\cong \mathbb{Z}_2$. The magnetic $\mathbb{Z}_2$ gauging on $\mathcal{T}^\vee$ is mirror to an electric $\mathbb{Z}_2$ gauging on $\mathcal{T}$.
%\begin{equation}
%    \mathcal{H}(\mathcal{T}_1)=\mathcal{C}(\mathcal{T}_1^\vee).
%\end{equation}
%\begin{equation}
%    \mathcal{C}(\mathcal{T}_1) =\mathcal{H}(\mathcal{T}_1^\vee).
%\end{equation}

The mirror of $\mathcal{T}^\vee_2$ is:
\begin{equation}
    \mathcal{T}_2=\raisebox{-0.5\height}{\begin{tikzpicture}
    \node[gauge,label=below:{$1$}] (0) at (0,0) {};
    \node[gauge,label=below:{$2$}] (1) at (1,0) {};
    \node[gauge,label=below:{$1$}] (2) at (2,0) {};
    \node[flavour,label=above:{$n$}] (1a) at (1,0.8) {};
    \draw (0)--(1)--(2) (1)--(1a);
    %\node at (0.75,0.3) {$n$};
    %\draw (0a) .. controls (4,1) .. (4);
    %\draw[transform canvas={yshift=2pt}] (0)--(1);
    %\draw[transform canvas={yshift=0pt}] (0)--(1);
    %\draw[transform canvas={yshift=-2pt}] (0)--(1);
    %\draw [thick,decoration={brace,mirror,raise=0.8cm},decorate] (1) -- (4) node [pos=0.5,anchor=north,yshift=-0.85cm] {$n-2$}; 
    %\draw (-.2,.6)--(0,0.4)--(.2,.6);
     %\node at (3.5,.5) {$k$};
     \end{tikzpicture}},
\label{eq_T2}
\end{equation}
which can be computed through a brane system \cite{Hanany:1996ie}.
%\begin{equation}
%    \mathcal{H}(\mathcal{T}_2)=\mathcal{C}(\mathcal{T}_2^\vee).
%\end{equation}
%\begin{equation}
%    \mathcal{C}(\mathcal{T}_2) =\mathcal{H}(\mathcal{T}_2^\vee).
%\end{equation}

The Higgs and Coulomb branches of $\mathcal{T}_1^\vee$ and $\mathcal{T}_2^\vee$ can be computed independently from their mirror descriptions, and the results are found to be in agreement.

\section{$A_1\cong C_1$}
\label{sec_5}

The case $n=2$ is exceptional and deserves separate treatment. In this limit the distinction between the $A$-type and $C$-type descriptions becomes degenerate, since $A_1\cong C_1$, and the general pattern discussed above does not apply directly.

For $n=2$, equation \eqref{eq_HTaffine} no longer holds. The Hilbert series of $\mathcal{H}(\mathcal{T})$ reads:
\begin{equation}
    \mathrm{HS}_\mathcal{H}(2)=\frac{1+2t^2+6t^4+2t^6+t^8}{(1-t^2)^2(1-t^4)^2}.
\end{equation}
This reproduces the Hilbert series of $a_2/\mathbb{Z}_2$ studied in Table 12 of \cite{Hanany:2023uzn}.

\begin{equation}
\mathcal{H}(\mathcal{T})\cong a_2/\mathbb{Z}_2.
\end{equation}
On the other hand, the affine closure of the minimal nilpotent orbit of $\mathfrak{sl}_2$ is isomorphic to $c_2$. Rather than fitting into the family of $\mathfrak{sl}_n$, it belongs to the $\mathfrak{sp}_n$ family~\cite{Fu:2026con}:

\begin{equation}
\overline{T^*\mathcal{O}^\mathrm{min}_{\mathfrak{sp}_n}}^{\mathrm{aff}}\cong c_{2n},\quad n\geq1,
\end{equation}
where $c_{2n}\cong \mathbb{C}^{4n}/\mathbb{Z}_2$.
Nevertheless, the theory $\mathcal{T}$ is still an interesting theory to study. Its Coulomb branch Hilbert series can be computed from the monopole formula:
\begin{equation}
    \mathrm{HS}_\mathcal{C}(2)=\frac{1+2t^2+6t^4+2t^6+t^8}{(1-t^2)^2(1-t^4)^2},
\end{equation}
which agrees with its Higgs branch Hilbert series. So this theory is self-mirror, i.e. it has identical Higgs and Coulomb branches:
\begin{equation}
    \mathcal{H}(\mathcal{T})\cong\mathcal{C}(\mathcal{T}).
\end{equation}
For the non-simply laced theory $\mathcal{T}^\vee$, the Hilbert series of its Coulomb branch can be computed using the monopole formula. One then finds:
\begin{equation}
\mathcal{C}(\mathcal{T}^\vee)\cong a_2/\mathbb{Z}_2.
\end{equation}
Due to the nature of isolated singularities, the inversion of $\mathcal{C}(\mathcal{T})$ requires prior knowledge of $\mathcal{H}(\mathcal{T})^\vee$, rendering the procedure essentially tautological. To establish $\mathcal{T}^\vee$ as self-mirror, a mathematical description of the Higgs branch for general non-simply laced theories is needed.
This would be an interesting direction for future work.

%\section{Outlook}
%We expect the inversion works for theories with each phase on Higgs/Coulomb branch has a well defined QFT description.
%The combinatorial condition to allow inversion for unitary quiver gauge theory is given in \cite{Bennett:2024loi}, where the non-trivial decoration leads to failure of inversion. More specific, on the Higgs branch of a gauge theory, the inversion is true if the Weyl group of the original gauge group does not normalise any possible residual gauge group. This is a potential physical interpretation of the special leaves defined in \cite{} and generalised in \cite{}.

\acknowledgments
We thank Antoine Bourget and Baohua Fu for useful discussions.
AH and DL are partially supported by STFC Consolidated Grants ST/T000791/1 and ST/X000575/1.

\appendix

\bibliographystyle{JHEP}
\bibliography{bibli.bib}

\end{document}